\documentclass[12pt]{book}

\usepackage{a4}
\usepackage{amsmath}
\usepackage{amssymb}
\usepackage{fancyhdr}
\usepackage{graphicx}
\usepackage{ifthen}
\usepackage{latexsym}
\usepackage{mathrsfs}
\usepackage{newlfont}
\newcounter{save}[chapter]

\newcommand{\3}{{\ss}}

\newcommand{\adjoint}[1]{{#1}^{\dagger}}

\newcommand{\bm}[1]{\mathbf{#1}}

\newcommand{\bra}[1]{\left\langle #1 \right|}

\newcommand{\braket}[3]{\ifthenelse{\equal{#2}{}}{\left\langle \left. #1 \right| #3 \right\rangle}{\left\langle #1 \left| #2 \right| #3 \right\rangle}}

\newcommand{\C}{\mathbb{C}}

\newcommand{\coco}[1]{{#1}^{\ast}}

\newcommand{\dd}[1]{d #1}

\newcommand{\ddd}[1]{d^3 #1}

\newcommand{\dddd}[1]{d^4 #1}

\newcommand{\effCo}{\lambda_{\mathrm{eff}}}

\renewcommand{\Im}{\mbox{Im}}

\newcommand{\intl}{\int\limits}

\newcommand{\ket}[1]{\left| #1 \right\rangle}

\newcommand{\kk}{\bm{k}_{\bm{n}}}
\newcommand{\pp}{\bm{p}_{\bm{m}}}
\newcommand{\qq}{\bm{q}_{\bm{l}}}

\newcommand{\pathdd}[1]{\mathscr{D} #1}

\newcommand{\R}{\mathbb{R}}

\renewcommand{\Re}{\mbox{Re}}

\newcommand{\sign}{\mbox{\rm sign}}

\newcommand{\tr}{\mbox{tr}}

\newenvironment{myeqnarray}[1]%
{\setcounter{save}{\value{equation}} \renewcommand{\theequation}{#1} \begin{eqnarray}}%
{\end{eqnarray}\setcounter{equation}{\value{save}}}

\newenvironment{myequation}[1]%
{\setcounter{save}{\value{equation}} \renewcommand{\theequation}{#1} \begin{equation}}%
{\end{equation}\setcounter{equation}{\value{save}}}

\begin{document}

\begin{titlepage}
  \centering
  \vspace*{\fill}
  {\LARGE\bf Faculty of Physics and Astronomy}
  \vfill
  {\Large\bf University of Heidelberg}
  \vfill
  \vfill
  \vfill
  \vfill
  \vfill
  \vfill
  {\bf Diploma thesis\\[0.5cm]
   in Physics\\[0.5cm]
   submitted by\\[0.5cm]
   Markus M. M\"uller\\[0.5cm]
   born in Saarlouis\\[0.5cm]
   2002}
\end{titlepage}
\thispagestyle{empty}

\begin{titlepage}
  \centering
  \vspace*{\fill}
  {\huge\bf Kinetic Equations\\from the\\Two-Particle-Irreducible $\bm{1/N}$-Expansion\\ \ }
  \vfill
  \vspace{1cm}
  \vfill
  \vfill
  {\bf This diploma thesis has been carried out by Markus M. M\"uller at the\\[0.5cm]
   Institute for Theoretical Physics\\[0.5cm]
   under the supervision of\\[0.5cm]
   Prof. Christof Wetterich}
\end{titlepage}
\thispagestyle{empty}

\cleardoublepage
\thispagestyle{empty}

\centerline{\large\bf Kinetic Equations from the}
\vspace{1ex}
\centerline{\large\bf Two-Particle-Irreducible $\bm{1/N}$-Expansion}
\vfill
\vfill
\centerline{\bf Abstract}
\vfill
{\small We present kinetic equations that describe the evolution of 
$O(N)$-symmetric real sca\-lar quantum fields out of thermal equilibrium 
in a systematic 
nonperturbative approximation scheme. This description starts from the 
$1/N$-expansion of the 2PI 
effective action to next-to-leading order, which includes scattering and
memory effects. From this starting point one is lead to evolution equations 
for the propagator, which are nonlocal in time. Numerical solutions showed
that the propagator depends only very smoothly on the center coordinates 
already after moderate times, and that correlations between earlier and 
later times are suppressed exponentially, which causes an effective memory 
loss. Exploiting these two observations, we combine a first order 
gradient expansion with a Wigner transformation to derive our kinetic 
equations, which are local in time, from the nonlocal evolution equations. 
In contrast to standard descriptions based on loop expansions, our kinetic 
equations remain valid even for nonperturbatively large fluctuations. 
Additionally, employing a quasi-particle approximation, we eventually arrive 
at a generalized Boltzmann equation.}
\vfill
\vfill
\vfill
\vfill
\vfill
\vfill
\centerline{\large\bf Kinetische Gleichungen aus der}
\vspace{1ex}
\centerline{\large\bf Zwei-Teilchen-Irreduziblen $\bm{1/N}$-Entwicklung}
\vfill
\vfill
\centerline{\bf Zusammenfassung}
\vfill
{\small Wir stellen kinetische Gleichungen vor, welche die zeitliche 
Entwicklung von $O(N)$-symme\-trischen reellen skalaren Quantenfeldern im 
thermischen Nichtgleichge\-wicht in einer systematischen nichtpertubativen 
N\"aherung beschreiben. Diese Beschreibung geht aus von der $1/N$-Entwicklung 
der 2PI effektiven Wirkung. Die Entwicklung bis einschlie\3lich 
n\"achstf\"uhrender Ordnung ber\"ucksichtigt nichttriviale Streuprozesse sowie 
das Erinnerungsverm\"ogen des Systems. Aus diesem Grund gelangt man zu 
Bewegungsgleichungen f\"ur den Propagator, welche nichtlokal in der Zeit sind.
Numerische L\"osungen dieser Glei\-chungen zeigten einerseits, da\3 der 
Propagator bereits nach m\"a\3ig gro\3en Zeiten nur noch sehr schwach von den 
Zentralkoordinaten abh\"angt, und andererseits, da\3 Korrelationen 
zwischen fr\"uheren und sp\"ateren Zeiten exponentiell ged\"ampft werden. 
Diese beiden Beobachtungen erlauben es uns, durch die Kombination einer 
Gradientenentwicklung mit einer Wignertransformation zu kinetischen 
Gleichungen zu gelangen, welche lokal in der Zeit sind. Im Gegensatz zu 
gew\"ohnlichen Beschreibungen, die auf einer Loop-Entwicklung beruhen, bleiben
unsere Gleichungen auch dann noch g\"ultig, wenn das betrachtete System 
nichtpertubativ gro\3e Fluktuationen zeigt. Die zus\"atzliche Anwendung einer
Quasiteilchenn\"aherung f\"uhrt uns letzt\-endlich zu einer verallgemeinerten 
Boltzmann\-gleichung.}

\clearpage
\thispagestyle{empty}
\cleardoublepage

\pagenumbering{roman}
\tableofcontents
\thispagestyle{empty}

\chapter*{Introduction and Overview}
\markboth{\em Introduction and Overview}{\em Introduction and Overview}
\addcontentsline{toc}{chapter}{Introduction and Overview}
\pagenumbering{arabic}
\thispagestyle{empty}

In recent years we have witnessed an enormous increase of interest in the 
dynamics of quantum fields out of equilibrium. Strong motivation in elementary
particle physics comes in particular from current and upcoming relativistic
heavy ion collision experiments at RHIC and LHC. In these experiments one aims 
at the examination of the quark-gluon plasma, which is produced in a state 
far from equilibrium. In contrast to equilibrium, nonequilibrium phenomena 
keep the information about the details of the initial conditions. Therefore, 
their understanding 
furnishes the only means to learn something about the earlier stages of a
collision. Further nonequilibrium phenomena can be found in cosmology, where 
one is interested for example in the inflationary dynamics 
of the early universe. A paradigm in this context is the phenomenon of 
parametric resonance in quantum field theory, which represents an important
building block for our understanding the (pre)heating of the early universe
after a period of inflation \cite{BeSe,BeMu}. Another very interesting problem
found in astroparticle physics is the generation of the observed 
matter-antimatter asymmetry of the universe. Indeed, already in 1967 Sakharov
noticed that this asymmetry can only be generated in a universe out of thermal
equilibrium \cite{Sa}. Furthermore, closely related to 
the description of nonequilibrium dynamics there appears the fundamental 
question of how macroscopically irreversible and dissipative behaviour can 
arise from microscopically time-reversal invariant laws of nature 
\cite{BeCo,Be1}. Or in other words, how does a closed system that is initially 
arbitrarily far away from equilibrium equilibrate?

There have been many different approaches to the description of 
nonequilibrium phenomena in quantum field theory. If one can guarantee that 
the coupling is small 
enough, one can use perturbative descriptions based on loop expansions
\cite{BeCo}. This approach has its limitations in the fact that the 
coupling is in general not constant, as for example in QCD, and may become 
too large to be an appropriate expansion parameter. Efforts to find a
nonperturbative, time-reflection invariant approach include the 
mean-field-type leading order large-$N$ approximation of the effective 
action \cite{CoHaKlMo}. However, this approximation neglects scattering 
and shows infinitely many conserved quantities not present in the 
underlying theory, which prohibit the approach to thermal equilibrium
\cite{Be1,BeWe1}. An attempt to go beyond mean-field was the $1/N$-expansion 
of the 1PI effective action to next-to-leading order \cite{BeWe1}. But 
this approximation can be secular (unbounded) in time, and therefore fails 
to give a controlled description for time evolution problems \cite{BeWe2}.
Computations in classical statistical field theory can be used as an 
approximate means to describe the evolution of quantum fields 
out of equilibrium. These computations are expected to yield quantitatively 
reliable results if the effective particle number density is sufficiently 
large, such that quantum fluctuations are dominated by statistical 
fluctuations. Of course they cannot describe the approach to quantum
thermal equilibrium \cite{AaBoWe1,AaBe2}.

Substantial progress was achieved recently in Ref.~\cite{Be1}, where it was 
shown that the $1/N$-expansion of the 2PI effective action to next-to-leading 
order furnishes a controlled nonperturbative description of the dynamics of 
$O( N )$-symmetric real scalar quantum fields far from equilibrium, which is 
based on first principles. This approach does not suffer from any of the 
previously mentioned restrictions. It holds even for systems containing 
strongly interacting particles far from equilibrium. In particular, it removes
spurious constants of motion encountered in leading order approximations and
permits the description of thermalization. As the effective action contains 
all information on the quantum fluctuations of the system under consideration, 
this approach is valid even for small particle numbers where quantum 
fluctuations are dominating. These features allow one to obtain quantitatively 
reliable results even for large times. 

The starting point for this description is the 2PI effective action defined on 
the closed Schwinger-Keldysh time contour 
\cite{BeCo,Schw,Ke,Cr,Da,ChSuHaYu,CaHu1}. The condition, that the 2PI effective
action be stationary with respect to variations of the propagator, leads via 
the Schwinger-Dyson equation to evolution equations for the spectral function 
and the symmetric propagator that are equivalent to the Kada\-noff-Baym 
equations \cite{Be1,Da,BaKa,BlIa}. 
These evolution equations take direct scattering into account, which leads to 
memory integrals. These integrals over time history make the evolution 
equations nonlocal in time.
\begin{figure}[tb]
  \centering
  \includegraphics[angle=90, width=9.5cm]{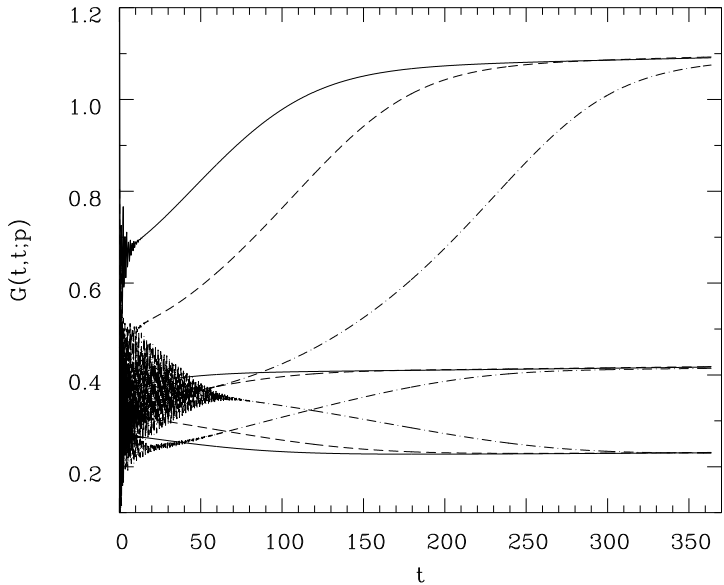}
  \begin{minipage}{12.5cm}
    \caption{\label{fig7}Evolution of the equal-time propagator for 
    three different momenta and three different initial conditions. We see 
    intermediate-time drifting and late-time thermalization. 
    Taken from \cite{BeCo}.}
  \end{minipage}
\end{figure}
Numerical solutions of these evolution equations (see Figs.~\ref{fig7} and
\ref{fig8}) show that the evolution of nonequilibrium quantum fields has 
three generic features \cite{Be1}:
\begin{enumerate}
  \item {\bf effective memory loss} \\
        (exponential suppression of unequal-time correlation functions, see 
        Fig.~\ref{fig8})
  \item {\bf intermediate-time drifting} \\
        (smooth, parametrically slow change of modes of the equal-time 
        correlation functions, see Fig.~\ref{fig7})
  \item {\bf late-time thermalization} \\
        (exponential approach to thermal equilibrium, see Fig.~\ref{fig7})
\end{enumerate}
\begin{figure}[tb]
  \centering
  \includegraphics[width=9.5cm]{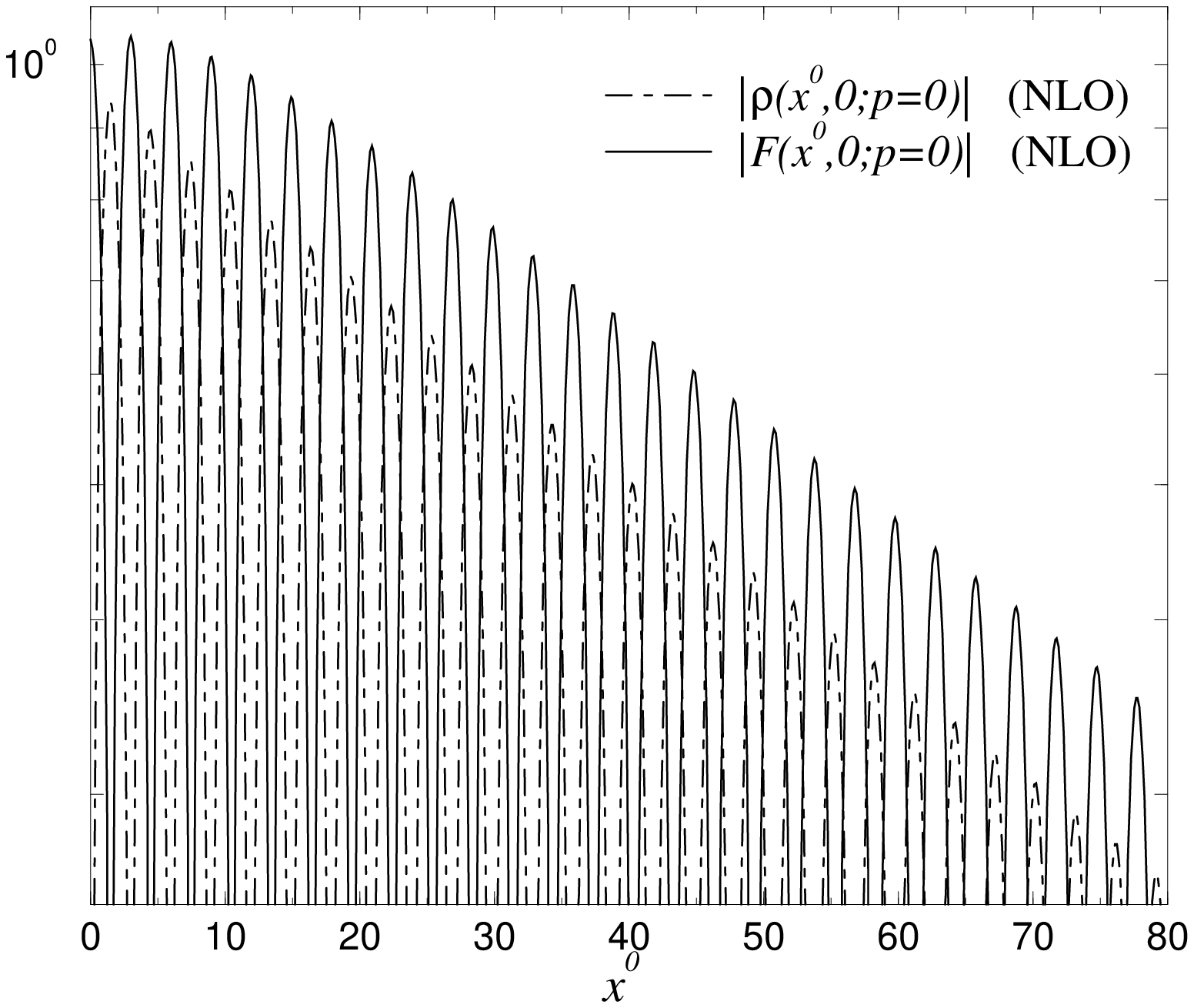}
  \begin{minipage}{12cm}
    \caption{\label{fig8}Logarithmic plot of the evolution of the 
    unequal-time spectral function and symmetric propagator. Correlations 
    between earlier and later times are exponentially suppressed, which 
    causes an effective memory loss. Taken from \cite{Be1}.}
  \end{minipage}
\end{figure}

While in equilibrium the propagator depends on the relative coordinates only, 
its additional dependence on the center coordinates is characteristic for its 
off-equilibrium behaviour. However, Fig.~\ref{fig7} shows that the 
evolution of the equal-time propagator in the intermediate-time drifting 
regime is very smooth. Furthermore, Fig.~\ref{fig8} shows that already for 
early times correlations between earlier and later times are exponentially 
suppressed, which means that there is an
effective memory loss. We will exploit these two properties of the 
propagator in order to describe the intermediate-time drifting and late-time
thermalization by kinetic equations which are local in time. We obtain these
kinetic equations from the nonlocal evolution equations by combining a first 
order gradient expansion with a Wigner transformation \cite{Da,BaKa,BlIa}. The 
application of the first order gradient expansion is justified by the smooth
dependence of the propagator on the center coordinates, and the effective 
memory loss allows us to send the limits of the memory integrals to
$\pm \infty$, which is a necessary condition for the Wigner transformation. 

The advantage of our kinetic equations is that their range of applicability 
is not limited by the validity of a loop expansion, but
instead is restricted only by the gradient expansion that we employed in 
their derivation. As we will show explicitly, they are even valid for 
nonperturbatively large fluctuations \cite{BeMu}. In such a situation there 
are contributions with leading order in the coupling from all 2PI loop orders,
which causes any loop expansion to break down. Compared to the full evolution
equations, in their range of applicability our kinetic equations have the 
advantage that they do not involve an integration over time history. This is
important for an efficient description of the late-time behaviour of 
nonequilibrium quantum fields. Apart from the description
of the intermediate-time drifting regime, as a further very important 
application these equations are expected to be valid near a second-order 
phase transition.

Additionally employing a quasi-particle approximation, we arrive at a 
generalized Boltzmann equation \cite{Da,BaKa,BlIa,La10}. Exactly as for the 
classical Boltzmann equation, its collision integral consists of terms 
representing the scattering of two quasi-particles which can be divided into 
gain and loss terms. However, in contrast to the classical Boltzmann equation
our generalized Boltzmann equation includes an effective mass and an effective 
coupling which arise from the contributions to the $1/N$-expansion of the 
2PI effective action to next-to-leading order. Due to the effective coupling
we expect our generalized Boltzmann equation to be valid even for large 
particle number densities.

In the first chapter we review briefly the description of quantum fields in 
and out of equilibrium using the imaginary-time path and the closed 
Schwinger-Keldysh time contour, respectively 
\cite{Schw,Ke,Cr,Da,ChSuHaYu,Ma,Be2,NiSe1,NiSe2}. 
In the second chapter we specialize to the description of 
far-from-equilibrium quantum fields using the $1/N$-expansion of the 2PI 
effective action to next-to-leading order. Here we review the derivation of 
the evolution equations that lead to Figs.~\ref{fig7} and \ref{fig8}
\cite{BeCo,Be1,CaHu1,AaAhBaBeSe}. 
In the third chapter we concentrate on the description of the 
intermediate-time drifting and late-time thermalization regimes and derive 
the kinetic equations and the generalized Boltzmann equation mentioned above. 
Finally, we give suggestions for a numerical implementation to solve the 
kinetic equations as well as the generalized Boltzmann equation
\cite{Da,CaHu1,BaKa,BlIa}.

\renewcommand{\theequation}{\arabic{chapter}.\arabic{section}.\arabic{equation}}

\chapter{Foundations}
\thispagestyle{empty}

In this chapter we review briefly some of the rather general foundations of our
considerations. After the introduction of our notation we consider quantum 
fields in equilibrium. The main motivation for this is Boltzmann's conjecture
that any closed system initially arbitrarily far away from equilibrium 
eventually equilibrates, which means that its late-time behaviour can 
effectively be described by an equilibrium ensemble average. In order to check 
whether our kinetic equations,
which we derive in chapter 3, contain a consistent description of the 
equilibrium, or whether the propagator has approached its equilibrium form 
to some given accuracy, we need to know some of the properties of the 
thermal propagator. These properties are reviewed in Sect.~1.2, using 
the imaginary time path introduced by Matsubara \cite{Ma,Be2}.

In contrast to equilibrium, for a nonequilibrated system the density 
operator depends on time, and the starting point to describe such a system 
changes dramatically. The main difference is that, while the equilibrium keeps 
no information on the details of the conditions for earlier times, the 
description of nonequilibrium phenomena is an initial value problem. However,
the techniques to describe such a system change only slightly. One has to 
switch to the closed Schwinger-Keldysh time contour
\cite{Schw,Ke,Cr,Da,ChSuHaYu}.

\section{Framework and Notation}

Throughout this work we use units where Planck's constant, the speed of light
and Boltzmann's constant are taken to be unity. We consider neutral spinless 
bosons of mass $m$, which can be described by a real scalar quantum field
\[ \Phi_a^I \left( x \right) = \frac{1}{\sqrt{2 \pi}^3} \int \ddd{p} \left[ \frac{1}{\sqrt{2 p^0}} \left( \exp \left( i p x \right) a_a \left( \bm{p} \right) + \exp \left( - i p x \right) \adjoint{a}_a \left( \bm{p} \right) \right) \right] \;. \]
The index $a \in \left\{ 1, \ldots, N \right\}$ enumerates the different 
particle species, $a_a$ and $\adjoint{a}_a$ are annihilation and creation 
operators respectively, and $p^0 = \sqrt{\bm{p}^2 + m^2}$. The superscript
$I$ indicates that $\Phi_a^I \left( x \right)$ is an {\em interaction picture}
quantum field. In the interaction picture time translations are generated by 
the free Hamiltonian $H_0$:
\begin{equation} \label{eq98}
  \Phi_a^I \left( x \right) = \exp \left( i x^0 H_0 \right) \Phi_a \left( \bm{x}, 0 \right) \exp \left( - i x^0 H_0 \right) \;.
\end{equation}
Quantum fields in the Heisenberg picture do not have any superscript at all.
Their time dependence is governed by the full Hamiltonian $H$:
\[ \Phi_a \left( x \right) = \exp \left( i x^0 H \right) \Phi_a \left( \bm{x}, 0 \right) \exp \left( - i x^0 H \right) \;. \]
The dynamics of the system under consideration are governed by the 
Lagrangian density
\[ \mathscr{L} \left( x \right) = \frac{1}{2} \Big( \partial_{\mu} \Phi_a \left( x \right) \Big) \Big( \partial^{\mu} \Phi_a \left( x \right) \Big) - \frac{1}{2} m^2 \Phi_a \left( x \right) \Phi_a \left( x \right) - \frac{\lambda}{4 ! N} \Big( \Phi_a \left( x \right) \Phi_a \left( x \right) \Big)^2 \;. \]
The plus sign of the kinetic term indicates that we use the metric where the
time-time component is positive. Of course, we use the summation convention
--- not only for the Lorentz indices, but also for the particle species 
indices. Being a bosonic quantum field, $\Phi_a \left( x \right)$ satisfies 
the equal-time commutation relation
\[ \left[ \Phi_a \left( \bm{x}, t \right); \Phi_b \left( \bm{y}, t \right) \right] = 0 \;. \]
Hence, for each $t$ there is a simultaneous eigenstate $\ket{ \varphi \left( t 
\right)}$ of the Heisenberg operators $\Phi_a \left( \bm{x}, t \right)$ with 
eigenvalues $\varphi_a \left( \bm{x} \right)$:
\[ \Phi_a \left( \bm{x}, t \right) \ket{\varphi \left( t \right)} = \varphi_a \left( \bm{x} \right) \ket{ \varphi \left( t \right)} \;. \]
The expectation value of the Heisenberg quantum field $\Phi_a \left( x 
\right)$ is denoted by
\[ \phi_a \left( x \right) = \left\langle \Phi_a \left( x \right) \right\rangle = \tr \left( \mathcal{D} \Phi_a \left( x \right) \right) \;. \]
The trace on the right hand side represents the ensemble average with respect
to the density operator $\mathcal{D}$, which describes the system. Accordingly,
the {\em connected two-point Green's function}, sometimes also called 
{\em connected propagator}, is given by:
\begin{equation} \label{eq63}
  G_{ab} \left( x, y \right) = \left\langle T_{\mathcal{T}} \left\{ \Phi_a \left( x \right) \Phi_b \left( y \right) \right\} \right\rangle - \phi_a \left( x \right) \phi_b \left( y \right) \;.
\end{equation}
The index $\mathcal{T}$ indicates that the time ordering follows a certain
time contour in the complex plane, i.e. the time coordinates are functions of 
the form
\[ z^0: \left[ 0; 1 \right] \to \C \;; \qquad u \mapsto z^0 \left( u \right) \;. \]
In the course of this work we will use different time contours, the form of 
which depends on the quantity under examination. In order to perform the time 
ordering explicitly we use $\Theta$ functions defined on the corresponding 
time contour by
\[ \Theta_{\mathcal{T}} \left( x^0 \left( u_x \right), y^0 \left( u_y \right) \right) = \theta \left( u_x - u_y \right) = \left\{ \begin{array}{ll} 1 & \mbox{, if } u_x > u_y \\ 0 & \mbox{, if } u_x < u_y \end{array} \right. \;. \]
Similarly, we can define contour $\delta$ functions:
\begin{eqnarray*}
        \delta_{\mathcal{T}} \left( x^0 \left( u_x \right), y^0 \left( u_y \right) \right) 
  & = & \frac{d}{d x^0} \Theta_{\mathcal{T}} \left( x^0 \left( u_x \right), y^0 \left( u_y \right) \right) \\
  & = & \frac{1}{\frac{d x^0}{d u_x}} \delta \left( u_x - u_y \right) \;.
\end{eqnarray*}
However, from now on we suppress the arguments of the time variables, and 
think of them as just living on the corresponding contour being ordered 
according to their emergence on the contour. Consequently, we use the 
following notation for integrals:
\[ \intl_{\mathcal{T}} \dddd{x} \left[ f \left( x \right) \right] = \int \ddd{x} \intl_0^1 \dd{u} \left[ f \left( \bm{x}, x^0 \left( u \right) \right) \left( \frac{d x^0 \left( u \right)}{du} \right) \right] \;. \]
Eventually, we also generalize the definition of the functional derivative 
according to
\[ \intl_{\mathcal{T}} \dddd{x} \left[ h \left( x \right) \frac{\delta F \left[ f \right]}{\delta f \left( x \right)} \right] = \lim_{\epsilon \to 0} \frac{1}{\epsilon} \left( F \left[ f + \epsilon h \right] - F \left[ f \right] \right) \;. \]
This definition leads immediately to
\[ \frac{\delta f \left( x \right)}{\delta f \left( y \right)} = \delta \left( \bm{x} - \bm{y} \right) \delta_{\mathcal{T}} \left( x^0, y^0 \right) \equiv \delta_{\mathcal{T}} \left( x - y \right) \;. \]
Defining the {\em two-point correlation functions} $G_{>, ab} \left( x, y 
\right)$ and $G_{<, ab} \left( x, y \right)$ by
\[ G_{>, ab} \left( x, y \right) \equiv \left\langle \Phi_a \left( x \right) \Phi_b \left( y \right) \right\rangle - \phi_a \left( x \right) \phi_b \left( y \right) \equiv G_{<, ba} \left( y, x \right) \;, \]
we can write the connected propagator in the form
\begin{equation} \label{eq1}
  G_{ab} \left( x, y \right) = \Theta_{\mathcal{T}} \left( x^0, y^0 \right) G_{>, ab} \left( x, y \right) + \Theta_{\mathcal{T}} \left( y^0, x^0 \right) G_{<, ab} \left( x, y \right) \;,
\end{equation}
a decomposition that will be of great use in subsequent chapters. While the
$\Theta$ functions appearing in Eq.~(\ref{eq1}) cause an inanalyticity of the
connected two-point Green's function with respect to its time arguments, the
correlation functions $G_{>, ab} \left( x, y \right)$ and $G_{<, ab} \left( x, 
y \right)$ are analytic functions of their time 
arguments. Another very useful property of the correlation functions is their 
so-called ``hermiticity'':
\begin{eqnarray*}
        \coco{G}_{>, ab} \left( x, y \right)
  & = & \coco{\left( \tr \left( \mathcal{D} \Phi_a \left( x \right) \Phi_b \left( y \right) \right) \right)} \\
  & = & \tr \left( \adjoint{\left( \mathcal{D} \Phi_a \left( x \right) \Phi_b \left( y \right) \right)} \right) \\
  & = & \tr \left( \Phi_b \left( y \right) \Phi_a \left( x \right) \mathcal{D} \right) \\
  & = & G_{>, ba} \left( y, x \right) \;.
\end{eqnarray*}
The same relation holds for $G_{<, ab} \left( x, y \right)$. The correlation 
functions $G_{>, ab} \left( x, y \right)$ and $G_{<, ab} \left( x, y \right)$ 
are also the constituents of some other frequently used quantities, such as 
the {\em retarded} and {\em advanced propagator}
\[ G_{R, ab} \left( x, y \right) = i \theta \left( x^0 - y^0 \right) \Big( G_{>, ab} \left( x, y \right) - G_{<, ab} \left( x, y \right) \Big) \;, \]
\[ G_{A, ab} \left( x, y \right) = - i \theta \left( y^0 - x^0 \right) \Big( G_{>, ab} \left( x, y \right) - G_{<, ab} \left( x, y \right) \Big) \;, \]
the {\em symmetric} or {\em statistical propagator}
\[ F_{ab} \left( x, y \right) = \frac{1}{2} \Big( G_{>, ab} \left( x, y \right) + G_{<, ab} \left( x, y \right) \Big) \;, \]
as well as the {\em spectral function}
\[ \varrho_{ab} \left( x, y \right) = G_{>, ab} \left( x, y \right) - G_{<, ab} \left( x, y \right) = \Big\langle \left[ \Phi_a \left( x \right); \Phi_b \left( y \right) \right] \Big\rangle \;. \]
We stress that we have used the usual $\theta$ function for real time 
arguments in the definitions of the retarded and the advanced propagator. 
Therefore, their definitions are meaningful only if $x^0$ and $y^0$ are real.
Because of the hermiticity of $G_{>, ab} \left( x, y \right)$ and $G_{<, ab} 
\left( x, y \right)$ the retarded and the advanced propagator are real 
functions. Furthermore, it follows immediately from the definition of 
$G_{>, ab} \left( x, y \right)$ and $G_{<, ab} \left( x, y \right)$ that
\[ G_{R, ab} \left( x, y \right) = G_{A, ba} \left( y, x \right) \;. \]

\section{\label{sect1}Thermal Quantum Field Theory}
\setcounter{equation}{0}

\begin{figure}[tb]
  \centering
  \includegraphics{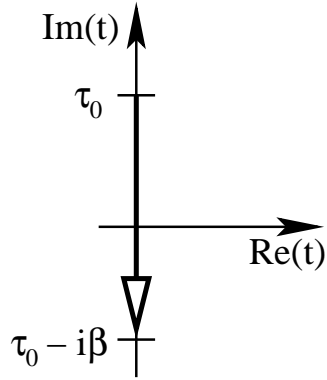} \\
  \begin{minipage}{10cm}
    \caption{\label{fig9}Imaginary time path, introduced by Matsubara 
    \cite{Ma}. Although we will use different values for $\tau_0$, we will 
    always use the same index $\mathcal{I}$.}
  \end{minipage}
\end{figure}

In this section we consider a closed system in equilibrium with 
time-independent Hamiltonian $H$ and temperature $T=\frac{1}{\beta}$. We will
work with the canonical density operator\footnote{Of course, there is particle
creation and particle annihilation in our system. Therefore, in equilibrium
the particle number will be such that the free energy is minimal:
\[ \frac{\partial F(T, V, N)}{\partial N} = 0 \]
The left hand side is just the definition of the chemical potential!}
\begin{equation} \label{eq42}
  \mathcal{D} = \frac{1}{Z} \exp \left( - \beta H \right) \;,
\end{equation}
where
\begin{equation} \label{eq62}
  Z = \tr \left( \exp \left( - \beta H \right) \right)
\end{equation}
is the {\em partition function}. The similarity between $\exp\left( - \beta H 
\right)$ and the time evolution operator $\exp \left( -i \left( t - t_0 
\right) H \right)$ suggests the definition of a new operator which contains 
these two operators as special cases. We set
\[ U \left( \tau \right) = \exp \left( - \tau H \right) \]
and allow for $\tau$ to take complex values. It is obvious that $U \left( \tau
\right)$ can be interpreted as an evolution operator for complex-valued times.
In order to compute the partition function we evaluate the trace using the 
complete set of eigenstates of the quantum field $\Phi_a \left( x \right)$ for 
$x^0=0$:
\[ Z = \int \prod_{a, \bm{x}} \dd{\varphi_a \left( \bm{x} \right)} \braket{\varphi \left( 0 \right)}{U \left( \beta \right)}{\varphi \left( 0 \right)} \;. \]
As indicated above, by interpreting $U \left( \tau \right)$ as an evolution
operator for complex times, we obtain the following path integral 
representation for the matrix element:
\[ \braket{\varphi \left( 0 \right)}{U \left( \beta \right)}{\varphi \left( 0 \right)} = \intl_{\begin{array}{c} \scriptstyle \varphi_a \left( \bm{x}, 0 \right) = \varphi_a \left( \bm{x} \right) \\ \scriptstyle \varphi_a \left( \bm{x}, -i \beta \right) = \varphi_a \left( \bm{x} \right) \end{array}} \prod_{a=1}^N \mathscr{D} \varphi_a \left( x \right) \left[ \exp \left( - S_E \left[ \varphi \right] \right) \right] \;. \]
Here $S_E$ denotes the Euclidean action which we obtain from the classical 
action by substituting $\tau = i x^0$.
\begin{eqnarray*}
        i S \left[ \varphi \right]
  & = & i \int \ddd{x} \intl_0^{- i \beta} \dd{x^0} \left[ \frac{1}{2} \left( \partial_{\mu} \varphi_a \right) \left( \partial^{\mu} \varphi_a \right) - \frac{1}{2} m^2 \varphi_a \varphi_a - \frac{\lambda}{4 ! N} \left( \varphi_a \varphi_a \right)^2 \right] \\
  & = & - \int \ddd{x} \intl_0^{\beta} \dd{\tau} \left[ \frac{1}{2} \delta^{\mu \nu} \left( \partial_{\mu} \varphi_a \right) \left( \partial_{\nu} \varphi_a \right) + \frac{1}{2} m^2 \varphi_a \varphi_a + \frac{\lambda}{4 ! N} \left( \varphi_a \varphi_a \right)^2 \right] \\
  & = & - S_E \left[ \varphi \right]
\end{eqnarray*}
Thus we arrive at the following path integral representation for the partition 
function:
\[ Z = \intl_{\varphi_a \left( \bm{x}, 0 \right) = \varphi_a \left( \bm{x}, - i \beta \right)} \prod_{a=1}^N \mathscr{D} \varphi_a \left( x \right) \left[ \exp \left( - S_E \left[ \varphi \right] \right) \right] \;. \]
By adding a source term to the Euclidean action, we turn the partition 
function into a generating functional:
\[ Z \left[ J \right] = \intl_{\varphi_a \left( \bm{x}, 0 \right) = \varphi_a \left( \bm{x}, - i \beta \right)} \prod_{a=1}^N \mathscr{D} \varphi_a \left( x \right) \left[ \exp \left( - S_E \left[ \varphi \right] + \int \ddd{x} \intl_0^{\beta} \dd{\tau} \left[ J_a \varphi_a \right] \right) \right] \;. \]
Using the standard procedure one can identify the $n$th functional derivative
of $Z \left[ J \right]$ with respect to $J$ with the expectation value of a
time-contour ordered product of $n$ quantum fields. The time contour that we 
have to use here is the imaginary time path $\mathcal{I}$ shown in 
Fig.~\ref{fig9}.
For example, we obtain for the two-point Green's function:
\begin{eqnarray*}
        \lefteqn{\frac{1}{Z} \left( \frac{\delta^2 Z \left[ J \right]}{\delta J_a \left( x \right) \delta J_b \left( y \right)} \right)_{J=0}} \; \\
  & = & \frac{1}{Z} \intl_{\varphi_c \left( \bm{z}, 0 \right) = \varphi_c \left( \bm{z}, - i \beta \right)} \prod_{c=1}^N \mathscr{D} \varphi_c \left( z \right) \left[ \varphi_a \left( x \right) \varphi_b \left( y \right) \exp \left( - S_E \left[ \varphi \right] \right) \right] \\
  & = & \left\langle T_{\mathcal{I}} \left\{ \Phi_a \left( x \right) \Phi_b \left( y \right) \right\} \right\rangle \\
  & = & \frac{1}{Z} \tr \Big( \exp \left( - \beta H \right) T_{\mathcal{I}} \left\{ \Phi_a \left( x \right) \Phi_b \left( y \right) \right\} \Big) \;.
\end{eqnarray*}
Correspondingly, the first derivative of the generating functional with
respect to the source gives the expectation value of the quantum field
$\Phi_a \left( x \right)$ and we obtain the connected thermal propagator as in
Eq.~(\ref{eq63}). Alternatively, we can use the generating functional for 
connected thermal $n$-point Green's functions
\[ W \left[ J \right] = \log \left( Z \left[ J \right] \right) \]
and obtain the connected thermal propagator as
\[ G_{ab} \left( x, y \right) = \left( \frac{\delta^2 W \left[ J \right]}{\delta J_a \left( x \right) \delta J_b \left( y \right)} \right)_{J=0} \;.\]
Because of the cyclic invariance of the trace we find that for $0 \leq \tau
\leq \beta$
\begin{eqnarray*}
        \lefteqn{\left\langle T_{\mathcal{I}} \left\{ \Phi_a \left( \bm{x}, - i \tau \right) \Phi_b \left( \bm{y}, 0 \right) \right\} \right\rangle} \; \\
  & = & \frac{1}{Z} \tr \Big( \exp \left( - \beta H \right) \exp \left( \tau H \right) \Phi_a \left( \bm{x}, 0 \right) \exp \left( - \tau H \right) \Phi_b \left( \bm{y}, 0 \right) \Big) \\
  & = & \frac{1}{Z} \tr \Big( \exp \left( - \beta H \right) \Phi_b \left( \bm{y}, 0 \right) \Phi_a \left( \bm{x}, -i \left( \tau - \beta \right) \right) \Big) \\
  & = & \left\langle T_{\mathcal{I}} \left\{ \Phi_a \left( \bm{x}, - i \left( \tau - \beta \right) \right) \Phi_b \left( \bm{y}, 0 \right) \right\} \right\rangle
\end{eqnarray*}
and
\[ \phi_a \left( \bm{x}, -i \tau \right) = \phi_a \left( \bm{x}, -i \left( \tau - \beta \right) \right) \;. \]
Hence, the connected thermal propagator satisfies the following periodicity 
condition:
\begin{equation} \label{eq91}
  G_{ab} \left( \bm{x}, -i \tau, \bm{y}, 0 \right) = G_{ab} \left( \bm{x}, -i \left( \tau - \beta \right), \bm{y}, 0 \right) \;.
\end{equation}
Instead of using a complete set of eigenstates of the quantum field $\Phi_a 
\left( x \right)$ for calculating $\tr \left( \mathcal{D} T_{\mathcal{I}} 
\left\{ \Phi_a \left( x \right) \Phi_b \left( y \right) \right\} \right)$ we 
also could have used a complete set of eigenstates of the Hamiltonian defined 
by
\[ H \ket{n} = E_n \ket{n} \;. \]
With their aid we obtain
\begin{eqnarray}
        \left\langle \Phi_a \left( x \right) \Phi_b \left( y \right) \right\rangle
  & = & \sum_{l,m,n} \braket{l}{\mathcal{D}}{n} \braket{n}{\Phi_a \left( x \right)}{m} \braket{m}{\Phi_b \left( y \right)}{l} \nonumber \\
  & = & \sum_{m,n} \frac{1}{Z} \braket{n}{\Phi_a \left( \bm{x}, 0 \right)}{m} \braket{m}{\Phi_b \left( \bm{y}, 0 \right)}{n} \label{eq65} \\
  &   & \times \exp \left( - i \left( x^0 -y^0 \right) E_m \right) \exp \left( - \left( \beta - i \left( x^0 - y^0 \right) \right) E_n \right) \nonumber
\end{eqnarray}
and
\begin{equation} \label{eq113}
  \phi_a \left( x \right) = \sum_n \exp \left( - \beta E_n \right) \braket{n}{\Phi_a \left( \bm{x}, 0 \right)}{n} \;.
\end{equation}
Eqs.~(\ref{eq65}) and (\ref{eq113}) show two properties of the thermal 
propagator. First, we see
that the time dependence of $G_{>, ab} \left( x, y \right)$ takes only $x^0 - 
y^0$ into account. The same also holds for $G_{<, ab} \left( x, y \right)$ and
we can conclude immediately that the thermal propagator is invariant under 
time translations\footnote{Although in our case $x^0$ and $y^0$ are purely 
imaginary quantities, this statement also holds for real times. In that case
one only has to switch to a different time contour \cite{Be2,NiSe1}.}. 
Additionally, the thermal propagator also is expected to be invariant under 
space translations, such that we finally arrive at a space-time translation 
invariant thermal propagator. Second, provided the matrix elements behave well 
enough, the convergence of the sum in Eq.~(\ref{eq65}) is controlled by the 
exponential functions and we find that $G_{>, ab} \left( x, y \right)$ is an 
analytic function of its time arguments in the domain
\begin{equation} \label{eq82}
  \left\{ x^0 - y^0 \in \C \Big| - \beta < \Im \left( x^0 - y^0 \right) < 0 \right\} \;.
\end{equation}
The first inequality is required by the second exponential function and vice
versa. By interchanging $x$ and $y$ in Eq.~(\ref{eq65}) $G_{<, ab} \left( x, y
\right)$ is found to be an analytic function of its time arguments in the 
domain
\begin{equation} \label{eq90}
  \left\{ x^0 - y^0 \in \C \Big| 0 < \Im \left( x^0 - y^0 \right) < \beta \right\} \;.
\end{equation}
In fact the domains (\ref{eq82}) and (\ref{eq90}) are promoted to be the 
domains of definition for $G_{>, ab} \left( x, y \right)$ and $G_{<, ab} 
\left( x, y \right)$, respectively. Knowing the domains of definition for 
$G_{>, ab} \left( x, y \right)$ and $G_{<, ab} \left( x, y \right)$ becomes 
important, if we look more closely at the periodicity condition (\ref{eq91}). 
We have
\[ G_{ab} \left( \bm{x}, -i \tau, \bm{y}, 0 \right) = G_{>, ab} \left( \bm{x}, -i \tau, \bm{y}, 0 \right) \;, \]
\[ G_{ab} \left( \bm{x}, -i \left( \tau - \beta \right), \bm{y}, 0 \right) = G_{<, ab} \left( \bm{x}, -i \left( \tau - \beta \right), \bm{y}, 0 \right) \]
and obtain the Kubo-Martin-Schwinger condition \cite{Ku,MaSchw}
\begin{equation} \label{eq92}
  \tilde{G}_{>, ab} \left( x - y \right) = \tilde{G}_{<, ab} \left( \bm{x} - \bm{y}, x^0 - y^0 + i \beta \right)
\end{equation}
by taking into account that in equilibrium the propagator is invariant under
space-time translations. The tilde indicates the use of relative coordinates 
as arguments. When we Fourier transform the Kubo-Martin-Schwinger condition
(\ref{eq92}) with respect to the relative coordinate $s = x - y$, we obtain
\begin{eqnarray}
        \tilde{G}_{>, ab} \left( k \right) 
  & = & \int \dddd{s} \left[ \exp \left( i k s \right) \tilde{G}_{>, ab} \left( s \right) \right] \nonumber \\
  & = & \exp \left( \beta k^0 \right) \tilde{G}_{<, ab} \left( k \right) \label{eq93} \;.
\end{eqnarray}
Using the Fourier transformed spectral function
\[ \tilde{\varrho}_{ab} \left( k \right) = \tilde{G}_{>, ab} \left( k \right) - \tilde{G}_{<, ab} \left( k \right) \]
we can write the Fourier transformed correlation functions in the form
\begin{equation} \label{eq94}
  \tilde{G}_{<, ab} \left( k \right) = \tilde{\varrho}_{ab} \left( k \right) n_B \left( k^0 \right) \;,
\end{equation}
\begin{equation} \label{eq95}
  \tilde{G}_{>, ab} \left( k \right) = \tilde{\varrho}_{ab} \left( k \right) \left( 1 + n_B \left( k^0 \right) \right) \;,
\end{equation}
where $n_B \left( k^0 \right)$ is the Bose-Einstein distribution function:
\[ n_B \left( k^0 \right) = \frac{1}{\exp \left( \beta k^0 \right) - 1} \;. \]
The main results of this section are the Fourier transformed 
Kubo-Martin-Schwin\-ger condition (\ref{eq93}) and Eqs.~(\ref{eq94}) and
(\ref{eq95}), which we will need in chapter 3. We have tried to arrive at these
results as quickly as possible without omitting important information and 
without
giving information not necessarily needed. Of course, there is much more to
say about thermal quantum field theory, e.g. one could want to calculate 
Green's functions with real time arguments. In this case one has to choose a
slightly different time contour as is explained in more detail for example in 
Refs.~\cite{Be2,NiSe1}.

\section[Quantum Fields out of Thermal Equilibrium]{\hspace{-2pt}Quantum Fields out of Thermal Equilibrium}
\setcounter{equation}{0}

The striking difference between equilibrium and nonequilibrium phenomena is 
that, while the canonical density operator (\ref{eq42}) is time-independent,
a density operator describing a system out of equilibrium depends on time,
\[ \mathcal{D} = \mathcal{D} \left( t \right) \]
and in general is known only for some initial time 
\[ t = t_{init} \equiv 0 \;. \]
Therefore, the description of nonequilibrium phenomena is an initial value 
problem. One starts with a set of initial $n$-point Green's functions given
by the initial density operator as
\begin{equation} \label{eq97}
  \tr \Big( \mathcal{D} \left( 0 \right) \Phi_{a_1} \left( \bm{x}_1, 0 \right) \ldots \Phi_{a_n} \left( \bm{x}_n, 0 \right) \Big)
\end{equation}
and follows their evolution in time. This will be our task for the rest of 
this work. In order to be able to describe general 
nonequilibrium initial conditions as well as initial thermal equilibrium, we
allow for the density operator to include mixed states, such that
\[ \tr \left( \mathcal{D}^2 \left( 0 \right) \right) < 1 \;. \]
In what follows we will restrict our considerations to spatially homogeneous
systems which can 
initially be described by a Gaussian density operator, i.e. the only 
non-vanishing initial correlations are completely determined by the one- and
two-point Green's functions and their first-order time derivatives at the
initial time. Hence, the set of initial $n$-point Green's functions we start 
with reads
\[ \phi_a \left( \bm{x}, 0 \right) = \tr \left( \mathcal{D} \left( 0 \right) \Phi_a \left( \bm{x}, 0 \right) \right) \;, \]
\[ \dot{\phi}_a \left( \bm{x}, 0 \right) = \tr \Big( \mathcal{D} \left( 0 \right) \left( \partial_{x^0} \Phi_a \left( x \right) \right) \Big)_{x^0 = 0} \;, \]
\[ G_{ab} \left( \bm{x}, 0, \bm{y}, 0 \right) = \tr \Big( \mathcal{D} \left( 0 \right) \Phi_a \left( \bm{x}, 0 \right) \Phi_b \left( \bm{y}, 0 \right) \Big) - \phi_a \left( \bm{x}, 0 \right) \phi_b \left( \bm{y}, 0 \right) \;, \]
\[ H_{ab} \left( \bm{x}, 0, \bm{y}, 0 \right) = \tr \Big( \mathcal{D} \left( 0 \right) \left( \partial_{x^0} \Phi_a \left( x \right) \right) \Phi_b \left( y \right) \Big)_{x^0 = y^0 = 0} - \dot{\phi}_a \left( \bm{x}, 0 \right) \phi_b \left( \bm{y}, 0 \right) \;, \]
\[ K_{ab} \left( \bm{x}, 0, \bm{y}, 0 \right) = \tr \Big( \mathcal{D} \left( 0 \right) \left( \partial_{x^0} \Phi_a \left( x \right) \right) \left( \partial_{y^0} \Phi_b \left( y \right) \right) \Big)_{x^0 = y^0 = 0} - \dot{\phi}_a \left( \bm{x}, 0 \right) \dot{\phi}_b \left( \bm{y}, 0 \right) \;. \]
We want to emphasize that the restriction on Gaussian initial conditions is
not an approximation at all. It only restricts the set of systems that can be
considered. The treatment of higher initial correlations imposes technical 
complications only, not fundamental ones \cite{BeCo}.

Next, we face the problem of calculating an expectation value of an arbitrary
composition of Heisenberg quantum fields for times later than the initial time.
In order to evaluate the trace (\ref{eq97}) in this more general case, we have
to take the evolution of each Heisenberg field with respect to the initial
time. Thus we obtain for the two-point correlation function:
\begin{eqnarray}
        \left\langle \Phi_a \left( x \right) \Phi_b \left( y \right) \right\rangle
  & = & \tr \Big( \mathcal{D} \left( 0 \right) \exp \left( i x^0 H \right) \Phi_a \left( \bm{x}, 0 \right) \exp \left( i \left( y^0 - x^0 \right) H \right) \nonumber \\
  &   & \qquad {} \times \Phi_b \left( \bm{y}, 0 \right) \exp \left( - i y^0 H \right) \Big) \nonumber \\
  & = & \tr \Big( \mathcal{D} \left( 0 \right) U \left( 0, x^0 \right) \Phi_a^I \left( x \right) U \left( x^0, y^0 \right) \Phi_b^I \left( y \right) U \left( y^0, 0 \right) \Big) \;. \qquad \quad \label{eq102}
\end{eqnarray}
In the second line we have used Eq.~(\ref{eq98}) and the interaction picture 
evolution operator
\begin{equation} \label{eq99}
  U \left( x^0, y^0 \right) = \exp \left( i x^0 H_0 \right) \exp \left( - i \left( x^0 - y^0 \right) H \right) \exp \left( - i y^0 H_0 \right) \;,
\end{equation}
where we have separated the full Hamiltonian $H$ into its free part $H_0$ and
the interaction $V$:
\[ H = H_0 + V \;. \]
Differentiating Eq.~(\ref{eq99}) with respect to $x^0$ gives the differential
equation
\begin{equation} \label{eq100}
  i \frac{d}{d x^0} U \left( x^0, y^0 \right) = V^I \left( x^0 \right) U \left( x^0, y^0 \right) \;,
\end{equation}
where 
\[ V^I \left( x^0 \right) = \exp \left( i x^0 H_0 \right) V \exp \left( - i x^0 H_0 \right) \;. \]
The interaction picture evolution operator (\ref{eq99}) is determined as the 
unique solution of the differential equation (\ref{eq100}) by the initial 
condition\footnote{This follows from the theorem of Picard-Lindel\"of on the 
existence and the uniqueness of solutions of ordinary differential equations.}
\begin{equation} \label{eq101}
  U \left( y^0, y^0 \right) = 1 \;.
\end{equation}
The differential equation (\ref{eq100}) together with the initial condition
(\ref{eq101}) is equivalent to the integral equation
\[ U \left( x^0, y^0 \right) = 1 - i \intl_{y^0}^{x^0} \dd{t} \left[ V^I \left( t \right) U \left( t, y^0 \right) \right] \;. \]
By iteration of this integral equation, we obtain an expansion for $U \left( 
x^0, y^0 \right)$ in powers of $V^I$:
\begin{eqnarray*}
        U \left( x^0, y^0 \right) 
  & = & 1 - i \intl_{y^0}^{x^0} \dd{t} \left[ V^I \left( t \right) \right] \\
  &   & {} + \left( -i \right)^2 \intl_{y^0}^{x^0} \dd{t_1} \intl_{y^0}^{t_1} \dd{t_2} \left[ V^I \left( t_1 \right) V^I \left( t_2 \right) \right] + \ldots \;.
\end{eqnarray*}
For $x^0 > y^0$ this expansion can be written in the form
\begin{eqnarray*}
        U \left( x^0, y^0 \right) 
  & = & 1 - i \intl_{y^0}^{x^0} \dd{t} \left[ V^I \left( t \right) \right] \\
  &   & {} + \frac{\left( -i \right)^2}{2} \intl_{y^0}^{x^0} \dd{t_1} \intl_{y^0}^{x^0} \dd{t_2} \left[ T \left\{ V^I \left( t_1 \right) V^I \left( t_2 \right) \right\} \right] + \ldots \\
  & = & T \exp \left( -i \intl_{y^0}^{x^0} \dd{t} \left[ V^I \left( t \right) \right] \right) \;,
\end{eqnarray*}
where $T$ denotes the usual time ordering operator along the real axis. On the
other hand, we obtain for $x^0 < y^0$
\begin{eqnarray*}
        U \left( x^0, y^0 \right) 
  & = & 1 + i \intl_{x^0}^{y^0} \dd{t} \left[ V^I \left( t \right) \right] \\
  &   & {} + \frac{i^2}{2} \intl_{x^0}^{y^0} \dd{t_1} \intl_{x^0}^{y^0} \dd{t_2} \left[ \tilde{T} \left\{ V^I \left( t_1 \right) V^I \left( t_2 \right) \right\} \right] + \ldots \\
  & = & \tilde{T} \exp \left( i \intl_{x^0}^{y^0} \dd{t} \left[ V^I \left( t \right) \right] \right) \;.
\end{eqnarray*}
Here $\tilde{T}$ denotes the antitemporal ordering along the real axis, i.e.
the interaction with the latest time argument is placed rightmost.

\begin{figure}[tb]
  \centering
  \includegraphics{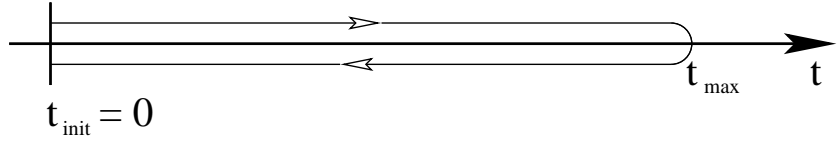}
  \begin{minipage}{12cm} 
    \caption{\label{fig10}Closed real-time path $\mathcal{C}$, introduced by 
    Schwinger \cite{Schw} and Keldysh \cite{Ke}.}
  \end{minipage}
\end{figure}

Now we return to the trace (\ref{eq102}) and find that all operators are 
sorted according to the order in which their time arguments appear on the 
time contour shown in Fig.~\ref{fig10} with 
\[ t_{max} = \max \left( x^0, y^0 \right) \;. \]
This is how the closed Schwinger-Keldysh time contour $\mathcal{C}$ comes
into play in a very natural way when one wants to describe quantum fields out
of equilibrium. Consequently, we have to define the $n$-point Green's 
functions on this time contour. For example the connected Schwinger-Keldysh
propagator is given by
\begin{equation} \label{eq96}
  G_{ab} \left( x, y \right) = \left\langle T_{\mathcal{C}} \left\{ \Phi_a \left( x \right) \Phi_b \left( y \right) \right\} \right\rangle - \phi_a \left( x \right) \phi_b \left( y \right) \;.
\end{equation}
With the aid of the density operator, the generating functional for these
Schwin\-ger-Keldysh Green's functions can be written in the form
\begin{eqnarray*}
        Z_{\mathcal{D}} \left[ J, K \right] 
  & = & \tr \Bigg( \mathcal{D} \left( 0 \right) T_{\mathcal{C}} \Big\{ \exp \Big( i \intl_{\mathcal{C}} \dddd{x} \left[ J_a \left( x \right) \Phi_a \left( x \right) \right] \\
  &   & {} + \frac{i}{2} \intl_{\mathcal{C}} \dddd{x} \intl_{\mathcal{C}} \dddd{y} \left[ \Phi_a \left( x \right) K_{ab} \left( x, y \right) \Phi_b \left( y \right) \right] \Big) \Big\} \Bigg) \;, \\
\end{eqnarray*}
where we have introduced the bilocal source $K$ for reasons that will become
clear in the next chapter, when we review the derivation of the 2PI effective
action. Exactly as
in the previous section, we evaluate the trace using the complete set of 
eigenstates of the quantum field $\Phi_a \left( x \right)$ for $x^0=0$:
\begin{eqnarray*}
        Z_{\mathcal{D}} \left[ J, K \right]
  & = & \int \prod_{c, \bm{x}} \dd{\varphi^{(1)}_c \left( \bm{x} \right)} \int \prod_{d, \bm{y}} \dd{\varphi^{(2)}_d \left( \bm{y} \right)} \Bigg[ \braket{\varphi^{(1)} \left( 0 \right)}{\mathcal{D} \left( 0 \right)}{\varphi^{(2)} \left( 0 \right)} \\
  &   & {} \times \bra{\varphi^{(2)} \left( 0 \right)} T_{\mathcal{C}} \Big\{ \exp \Big( i \intl_{\mathcal{C}} \dddd{x} \left[ J_a \left( x \right) \Phi_a \left( x \right) \right] \\
  &   & {} + \frac{i}{2} \intl_{\mathcal{C}} \dddd{x} \intl_{\mathcal{C}} \dddd{y} \left[ \Phi_a \left( x \right) K_{ab} \left( x, y \right) \Phi_b \left( y \right) \right] \Big) \Big\} \ket{\varphi^{(1)} \left( 0 \right)} \Bigg] \;.
\end{eqnarray*}
The second matrix element is given by the following path integral:
\begin{eqnarray*}
  \intl_{\begin{array}{c} \scriptstyle \varphi_c \left( \bm{x}, \overrightarrow{0} \right) = \varphi^{(1)}_c \left( \bm{x} \right) \\ \scriptstyle \varphi_c \left( \bm{x}, \overleftarrow{0} \right) = \varphi^{(2)}_c \left( \bm{x} \right) \end{array}} \prod_{c=1}^N \mathscr{D} \varphi_c \left( x \right) \Bigg[ \exp \Big( i \intl_{\mathcal{C}} \dddd{x} \left[ \mathscr{L} \left( x \right) + J_a \left( x \right) \varphi_a \left( x \right) \right] && \\
  \qquad \qquad {} + \frac{i}{2} \intl_{\mathcal{C}} \dddd{x} \intl_{\mathcal{C}} \dddd{y} \left[ \varphi_a \left( x \right) K_{ab} \left( x, y \right) \varphi_b \left( y \right) \right] \Big) \Bigg] \;. &&
\end{eqnarray*}
The matrix element of the density operator can be written in the form
\[ \braket{\varphi^{(1)} \left( 0 \right)}{\mathcal{D} \left( 0 \right)}{\varphi^{(2)} \left( 0 \right)} = \exp \left( i F \left[ \varphi \right] \right) \;, \]
where $F$ can be expanded in the functional sense according to
\begin{eqnarray}
        F \left[ \varphi \right]
  & = & \alpha^{(0)} + \intl_{\mathcal{C}} \dddd{x} \left[ \alpha^{(1)}_a \left( x \right) \varphi_a \left( x \right) \right] \nonumber \\
  &   & {} + \frac{1}{2} \intl_{\mathcal{C}} \dddd{x} \intl_{\mathcal{C}} \dddd{y} \left[ \alpha^{(2)}_{ab} \left( x, y \right) \varphi_a \left( x \right) \varphi_b \left( y \right) \right] \label{eq110} \\
  &   & {} + \frac{1}{3 !} \intl_{\mathcal{C}} \dddd{x} \intl_{\mathcal{C}} \dddd{y} \intl_{\mathcal{C}} \dddd{z} \left[ \alpha^{(3)}_{abc} \left( x, y, z \right) \varphi_a \left( x \right) \varphi_b \left( y \right) \varphi_c \left( z \right) \right] \nonumber \\
  &   & {} + \ldots \;. \nonumber
\end{eqnarray}
The values of $\varphi_a \left( x \right)$ at the beginning and the end
of the time contour are given by $\varphi^{(1)}_a \left( \bm{x} \right)$
and $\varphi^{(2)}_a \left( \bm{x} \right)$, respectively.
Of course, the sequence of the coefficients $\alpha^{(j)}$ contains as much 
information as the original density matrix. So we can rewrite the above trace
as a functional of infinitely many nonlocal sources. However, at the beginning 
of this section we restricted our considerations to Gaussian initial 
conditions. This means for the sources $\alpha^{(j)}$ that
\[ \alpha^{(j)}_{a_1, \ldots, a_j} \left( x_1, \ldots, x_j \right) = 0 \qquad \left( \forall j \geq 3 \right) \;. \]
Even the sources $\alpha^{(1)}$ and $\alpha^{(2)}$ are non-vanishing only at 
the initial time $t_{init} = 0$, such that we can absorb
them into the external sources $J$ and $K$. Anyway, we can only compute ratios 
of path integrals. In such a ratio the constant factor caused by 
$\alpha^{(0)}$ cancels, and therefore, we can cancel $\alpha^{(0)}$ 
from the functional (\ref{eq110}). Thus, the generating functional for 
Schwinger-Keldysh Green's functions is given by the path integral
\begin{eqnarray}
        Z_{\mathcal{D}} \left[ J, K \right] 
  & = & \int \prod_{c=1}^N \pathdd{\varphi_c \left( x \right)} \left[ \exp \left( i \intl_{\mathcal{C}} \dddd{x} \left[ \mathscr{L} \left( x \right) + J_a \left( x \right) \varphi_a \left( x \right) \right] \right. \right. \nonumber \\
  &   & \qquad \left. \left. {} + \frac{i}{2} \intl_{\mathcal{C}} \dddd{x} \intl_{\mathcal{C}} \dddd{y} \left[ \varphi_a \left( x \right) K_{ab} \left( x, y \right) \varphi_b \left( y \right) \right] \right) \right] \label{eq109}
\end{eqnarray}
and the expectation values needed for the calculation of the connected 
Schwin\-ger-Keldysh propagator (\ref{eq96}) are given by
\[ \left\langle T_{\mathcal{C}} \left\{ \Phi_a \left( x \right) \Phi_b \left( y \right) \right\} \right\rangle = \frac{\left( - i \right)^2}{Z_{\mathcal{D}} \left[ 0, 0 \right]} \left( \frac{\delta^2 Z_{\mathcal{D}} \left[ J, K \right]}{\delta J_a \left( x \right) \delta J_b \left( y \right)} \right)_{J = K = 0} \]
and
\[ \phi_a \left( x \right) = \frac{- i}{Z_{\mathcal{D}} \left[ 0, 0 \right]} \left( \frac{\delta Z_{\mathcal{D}} \left[ J, K \right]}{\delta J_a \left( x \right)} \right)_{J = K = 0} \;. \]
It is important to note that the closed Schwinger-Keldysh time contour 
emerges necessarily whenever one wants to treat nonequilibrium phenomena, 
regardless of the special circumstances (elementary particle physics,
description of conductors and semi-conductor devices, cosmology, \ldots). 

In the next chapter we will start from the generating functional (\ref{eq109}) 
and derive the two-particle irreducible effective action defined on the 
closed Schwinger-Keldysh time contour. After having presented the 
$1/N$-expansion of the 2PI effective action, we then are going to derive
equations which determine the evolution of the Schwinger-Keldysh propagator.

\chapter{Nonlocal Dynamics of Quantum Fields out of Equilibrium}
\thispagestyle{empty}

In this chapter we continue our considerations on quantum fields out of 
thermal equilibrium. Starting from the generating functional for 
Schwinger-Keldysh Green's functions we derive the two-particle irreducible
effective action defined on the closed Schwinger-Keldysh time contour. We 
review the $1/N$-expansion of the 2PI effective action to next-to-leading 
order as well as the derivation of the corresponding evolution equations for
the propagator, which lead to Figs.~\ref{fig7} and \ref{fig8} in the 
introduction. As we already mentioned in the introduction, these evolution 
equations furnish a controlled nonperturbative description of the 
time-reversal invariant dynamics of quantum fields far from equilibrium.

\section{2PI Effective Action}
\setcounter{equation}{0}

The {\em effective action} was introduced perturbatively by Jeffrey Goldstone,
Abdus Salam and Steven Weinberg in 1962 \cite{GoSaWe}. They defined it
as the sum over all one-particle irreducible\footnote{A diagram is called 
$n$-particle-irreducible ($n$PI), if it cannot be disconnected by cutting
through any $n$ internal lines.} diagrams. In the following year Bryce S. 
DeWitt gave the nonperturbative definition \cite{DeWi}. The physical 
significance of the effective action arises for two reasons. First, it is the 
generating functional for 1PI $n$-point Green's functions, and second it 
provides an equation of motion for the expectation value of the quantum field, 
which includes quantum corrections. Later on, this form of the effective 
action was called 1PI effective action in order to distinguish it from higher 
effective actions. 

In this section we are going to summarize briefly the basic facts about the 
2PI effective action $\Gamma \left[ \phi, G \right]$. In analogy to what we
said above, it is the generating functional for 2PI $n$-point Green's 
functions expressed in terms of the connected propagator $G_{ab} \left( x, y 
\right)$. This means that the diagrammatic expansion of for example
\[ \left( \frac{\delta^2 \Gamma \left[ \phi, G \right]}{\delta \phi_a \left( x \right) \delta \phi_b \left( y \right)} \right)_{\phi = 0} \]
contains only two-particle irreducible diagrams, where internal lines 
represent the connected propagator. Apart from that, the 2PI effective action
provides equations of
motion for the field expectation value $\phi_a \left( x \right)$ and the 
connected propagator $G_{ab} \left( x, y \right)$, which take quantum 
corrections into account.

The starting point for the derivation of an explicit expression for the 2PI 
effective action is the path integral representation
of the generating functional for Schwinger-Keldysh Green's functions 
(\ref{eq109}), namely:
\begin{eqnarray*}
        Z \left[ J, K \right] 
  & = & \int \prod_{c=1}^N \pathdd{\varphi_c \left( x \right)} \left[ \exp \left( i \intl_{\mathcal{C}} \dddd{x} \left[ \mathscr{L} \left( x \right) + J_a \left( x \right) \varphi_a \left( x \right) \right] \right. \right. \\
  &   & \qquad \left. \left. {} + \frac{i}{2} \intl_{\mathcal{C}} \dddd{x} \intl_{\mathcal{C}} \dddd{y} \left[ \varphi_a \left( x \right) K_{ab} \left( x, y \right) \varphi_b \left( y \right) \right] \right) \right] \;.
\end{eqnarray*}
The connected $n$-point Green's functions are obtained by taking the $n$-th
functional derivative of 
\[ W \left[ J, K \right] = - i \log \left( Z \left[ J, K \right] \right) \]
with respect to $J$. In particular, for $n=1$ we obtain the expectation value 
of the field $\Phi_a \left( x \right)$:
\begin{equation} \label{eq2}
  \left( \frac{\delta W \left[ J, K \right]}{\delta J_a \left( x \right)} \right)_{J=K=0} = \phi_a \left( x \right) \;.
\end{equation}
However, the first derivative of $W \left[ J, K \right]$ with respect to $K$
gives the full 2-point Green's function, including disconnected diagrams:
\begin{equation} \label{eq3}
  \left( \frac{\delta W \left[ J, K \right]}{\delta K_{ab} \left( x, y \right)} \right)_{J=K=0} = \frac{1}{2} \left( G_{ab} \left( x, y \right) + \phi_a \left( x \right) \phi_b \left( y \right) \right) \;.
\end{equation}
Using Eqs.~(\ref{eq2}) and (\ref{eq3}), one can express $J$ and $K$ in terms 
of $\phi$ and $G$. The 2PI effective action $\Gamma \left[ \phi, 
G \right]$ is the double Legendre transform of $W \left[ J, K 
\right]$:
\begin{eqnarray*}
        \Gamma \left[ \phi, G \right]
  & = & W \left[ J, K \right] - \intl_{\mathcal{C}} \dddd{x} \left[ J_a \left( x \right) \phi_a \left( x \right) \right] \\
  &   & {} - \frac{1}{2} \intl_{\mathcal{C}} \dddd{x} \intl_{\mathcal{C}} \dddd{y} \Big[ G_{ab} \left( x, y \right) K_{ab} \left( x, y \right) \\
  &   & {} + \phi_a \left( x \right) K_{ab} \left( x, y \right) \phi_b \left( y \right) \Big]
\end{eqnarray*}
From this, we see that
\[ \frac{\delta \Gamma \left[ \phi, G \right]}{\delta \phi_a \left( x \right)} = - J_a \left( x \right) - \intl_{\mathcal{C}} \dddd{y} \left[ K_{ab} \left( x, y \right) \phi_b \left( y \right) \right] \]
and
\[ \frac{\delta \Gamma \left[ \phi, G \right]}{\delta G_{ab} \left( x, y \right)} = - \frac{1}{2} K_{ab} \left( x, y \right) \;. \]
In the case of vanishing external sources $J$ and $K$ the equations of motion 
for $\phi$ and $G$ read
\begin{equation} \label{eq5}
  \frac{\delta \Gamma \left[ \phi, G \right]}{\delta \phi_a \left( x \right)} = 0 \;,
\end{equation}
\begin{equation} \label{eq6}
  \frac{\delta \Gamma \left[ \phi, G \right]}{\delta G_{ab} \left( x, y \right)} = 0 \;.
\end{equation}
The usual 1PI effective action $\Gamma \left[ \phi \right]$ is obtained by
Legendre transforming the generating functional for connected $n$-point 
Green's functions with a vanishing bilocal source. Having this in mind, one 
sees the connection between $\Gamma \left[ \phi \right]$ and $\Gamma \left[ 
\phi, G \right]$:
\begin{equation} \label{eq7}
  \Gamma \left[ \phi \right] = \Gamma \left[ \phi, G_{stat} \right] \;,
\end{equation}
where $G_{stat}$ satisfies the stationarity condition (\ref{eq6}):
\[ \left( \frac{\delta \Gamma \left[ \phi, G \right]}{\delta G_{ab} \left( x, y \right)} \right)_{G = G_{stat}} = 0 \;. \]
In order to solve Eqs.~(\ref{eq5}) and (\ref{eq6}) it is necessary to find an 
explicit expression for $\Gamma \left[ \phi, G \right]$. John M. 
Cornwall, R. Jackiw and E. Tomboulis provided the following parametrization 
in Ref.~\cite{CoJaTo}:
\begin{eqnarray}
        \Gamma \left[ \phi, G \right] 
  & = & S \left[ \phi \right] + \frac{i}{2} \tr_{\mathcal{C}} \left[ \log \left[ G^{-1} \right] \right] + \frac{i}{2} \tr_{\mathcal{C}} \left[ G_0^{-1} G \right] \label{eq8} \\
  &   & + \Gamma_2 \left[ \phi, G \right] + const \nonumber \;.
\end{eqnarray}
Here the squared brackets indicate that the traces, the logarithm and the 
product $G_0^{-1} G$ have to be evaluated in the functional sense. $S \left[ 
\phi \right]$ is the classical action defined on the closed Schwinger-Keldysh
time contour and
\begin{eqnarray}
        i G_{0, ab}^{-1} \left( x, y \right) 
  & = & \frac{\delta^2 S \left[ \phi \right]}{\delta \phi_a \left( x \right) \delta \phi_b \left( y \right)} \nonumber \\
  & = & \left( - \Box_x - m^2 - \frac{\lambda}{3 ! N} \phi_d \left( x \right) \phi_d \left( x \right) \right) \delta_{ab} \delta_{\mathcal{C}} \left( x - y \right) \label{eq72} \\
  &   & {} - \frac{\lambda}{3 N} \phi_a \left( x \right) \phi_b \left( x \right) \delta_{\mathcal{C}} \left( x - y \right) \nonumber
\end{eqnarray}
is the inverse classical propagator. The constant can be chosen such that 
Eq. (\ref{eq7}) is indeed satisfied. To compute $\Gamma_2$ we define a new 
field which has vanishing expectation value:
\[ \Phi'_a \left( x \right) = \Phi_a \left( x \right) - \phi_a \left( x \right) \;. \]
Then we rewrite the classical action in terms of $\Phi_a' \left( x \right)$ 
and $\phi_a \left( x \right)$:
\[ S \left[ \Phi \right] = S \left[ \Phi' + \phi \right] \;. \]
Apart from linear and quadratic terms, the action on the right hand side 
includes terms cubic and quartic in the field $\Phi'_a \left( x \right)$. 
$\Gamma_2 \left[ \phi, G \right]$ is the sum of all two-particle irreducible 
vacuum graphs with vertices determined by those cubic and quartic terms, while 
internal lines are set equal to the connected propagator $G_{ab} \left( x, y 
\right)$.

\section[$1/N$-Expansion of the 2PI Effective Action]{$\bm{1/N}$-Expansion of the 2PI Effective Action}
\setcounter{equation}{0}

In a free field theory $\Gamma_2$ vanishes and one can compute the 2PI 
effective action exactly. However, this is not possible for a theory that
contains interactions. When this is the case, one has to look for an 
appropriate approximation scheme. In this section, we review briefly the 
$1/N$-expansion of $\Gamma_2$ to next-to-leading order (NLO). Here, we 
restrict ourselves to the presentation of the results only. Their derivation 
is described in more detail in Refs.~\cite{Be1,AaAhBaBeSe}.

\begin{figure}[b]
  \centering
  \includegraphics{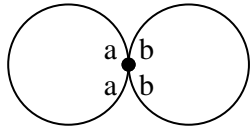}
  \caption{\label{fig1}Leading order contribution to $\Gamma_2$.}
\end{figure}

\begin{figure}[tb]
  \centering
  \includegraphics{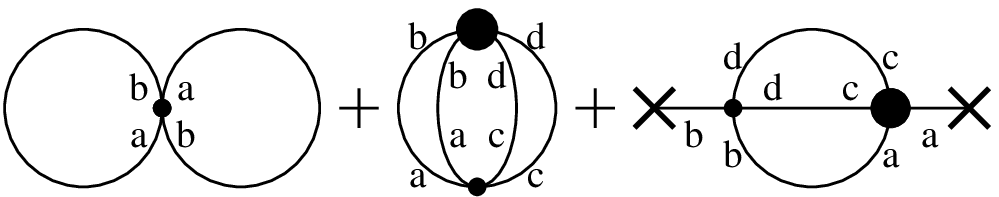} \\
  \begin{minipage}{12cm}
    \caption{\label{fig5}Next-to-leading order contribution to $\Gamma_2$. 
    The bulky vertices are defined in Fig.~\ref{fig6}. Crosses indicate an
    expectation value of the original quantum field.}
  \end{minipage} \\[1cm]
  \includegraphics{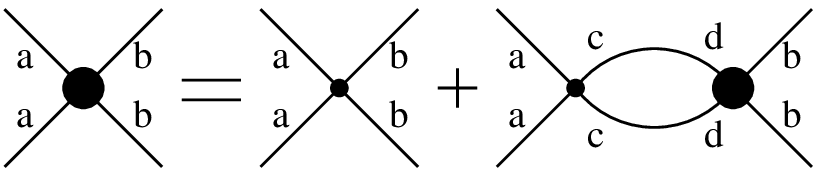} \\
  \begin{minipage}{10cm}
     \caption{\label{fig6} Resummation scheme for the next-to-leading order 
     contribution to $\Gamma_2 \left[ \phi, G \right]$.}
  \end{minipage}
\end{figure}

To obtain the $1/N$-expansion of $\Gamma_2 \left[ \phi, G \right]$ we have to 
sort all the diagrams contributing to $\Gamma_2 \left[ \phi, G \right]$ with
respect to the powers of their $1/N$ factors, beginning with the lowest powers
and proceeding to higher powers. We take all diagrams into account that 
contribute to next-to-leading order, i.e. we write $\Gamma_2 \left[ \phi, 
G \right]$ in the form
\begin{equation} \label{eq9}
  \Gamma_2 \left[ \phi, G \right] = \Gamma_2^{(LO)} \left[ \phi, G \right] + \Gamma_2^{(NLO)} \left[ \phi, G \right] + \ldots
\end{equation}
and neglect terms beyond next-to-leading order in $1/N$. Fig.~\ref{fig1} 
shows the only diagram that contributes to leading order.
Hence:
\[ \Gamma_2^{(LO)} \left[ \phi, G \right] = - \frac{\lambda}{4! N} \intl_{\mathcal{C}} \dddd{x} \left[ G_{aa} \left( x, x \right) G_{bb} \left( x, x \right) \right] \;. \]
The next-to-leading order contribution is shown in Fig.~\ref{fig5}.
A bulky vertex indicates an infinite series of diagrams, where each diagram 
has one vertex and one loop more than its predecessor. The corresponding 
recursively defined resummation scheme is shown in Fig.~\ref{fig6}. We obtain
\begin{eqnarray*}
        \Gamma_2^{(NLO)} \left[ \phi, G \right] 
  & = & \frac{i}{2} \tr_{\mathcal{C}} \left[ \log \left[ B \right] \right] \\
  &   & {} + \frac{i \lambda}{6 N} \intl_{\mathcal{C}} \dddd{x} \intl_{\mathcal{C}} \dddd{y} \left[ I \left( x, y \right) \phi_a \left( x \right) G_{ab} \left( x, y \right) \phi_b \left( y \right) \right] \;.
\end{eqnarray*}
Here 
\begin{equation} \label{eq111}
  I \left( x, y \right) = \frac{\lambda}{6 N} G_{ab} \left( x, y \right) G_{ab} \left( x, y \right) - \frac{i \lambda}{6 N} \intl_{\mathcal{C}} \dddd{z} \left[ I \left( x, z \right) G_{ab} \left( z, y \right) G_{ab} \left( z, y \right) \right]
\end{equation}
resums the chain of bubbles caused by one bulky vertex. For the diagrams that 
contain exclusively four-point vertices this resummation gives the trace 
over the functional series of the logarithm of
\begin{equation} \label{eq10}
  B \left( x, y \right) = \delta_{\mathcal{C}} \left( x - y \right) + \frac{i \lambda}{6 N} G_{ab} \left( x, y \right) G_{ab} \left( x, y \right) \;.
\end{equation}
In the symmetric regime ($\phi=0$) all diagrams that contain three-point 
vertices vanish and only the trace remains as the NLO contribution to 
$\Gamma_2 \left[ \phi, G \right]$. 

Eventually, we would like to illuminate the distinction in the resummation 
between the 2PI $1/N$-expansion and a perturbative expansion. Altogether, 
there are four diagrams that contribute to $\Gamma_2 \left[ \phi, G \right]$ 
to order $\lambda^2$, but only two of them\footnote{These are the second and
the third diagram shown in Fig.~\ref{fig5} with, according to Fig.~\ref{fig6},
the bulky vertex replaced by the classical one.} contribute to the 2PI 
$1/N$-expansion to next-to-leading order. Rotating one of the two vertices 
in each of these two diagrams by $90^{\circ}$ respectively, one obtains 
the remaining two diagrams, which contribute at NNLO in $1/N$. Similar 
considerations hold for higher loop diagrams. At each order in the coupling 
there are only two diagrams contributing to the 2PI $1/N$-expansion to 
next-to-leading order.

\section{\label{sect2}Evolution Equations}
\setcounter{equation}{0}

\begin{figure}[t]
  \centering
  \includegraphics{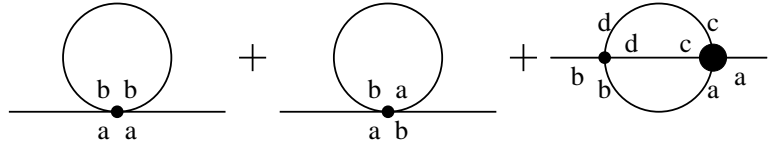}
  \begin{minipage}{12cm}
     \caption{\label{fig4}Diagrammatic expansion of the self energy in the
     symmetric regime. The bulky vertex represents the resummation that has 
     been introduced in Fig.~\ref{fig6}. Vertices with external lines are 
     linked to space time variables which must not be integrated over.}
  \end{minipage}
\end{figure}

From here on, for simplicity we restrict our considerations to the symmetric 
regime, where
\[ \phi_a \left( x \right) = 0 \;. \]
The starting point for the derivation of the evolution equations for the 
spectral function and the symmetric propagator is the equation of motion 
(\ref{eq6}). With the parametrization (\ref{eq8}) it reads
\begin{equation} \label{eq43}
  G_{ab}^{-1} \left( x, y \right) = G_{0, ab}^{-1} \left( x, y \right) - 2 i \frac{\delta \Gamma_2 \left[ G \right]}{\delta G_{ab} \left( x, y \right)} \;,
\end{equation}
which is nothing but the exact Schwinger-Dyson equation for the propagator
where the proper self energy is given by
\begin{equation} \label{eq44}
  \Sigma_{ab} \left( x, y \right) = 2 i \frac{\delta \Gamma_2 \left[ G \right]}{\delta G_{ab} \left( x, y \right)} \;.
\end{equation}
When we insert the inverse classical propagator (\ref{eq72}) (here for $\phi 
= 0$) as well as Eq.~(\ref{eq44}) the Schwinger-Dyson equation (\ref{eq43}) 
takes the form
\begin{equation} \label{eq73}
  G_{ab}^{-1} \left( x, y \right) = - \left( \Box_x + m^2 \right) \delta_{ab} \delta_{\mathcal{C}} \left( x - y \right) - \Sigma_{ab} \left( x, y \right) \;.
\end{equation}
At next-to-leading order in the $1/N$-expansion of the 2PI effective action
Eq. (\ref{eq44}) yields for the proper self energy:
\begin{equation} \label{eq13}
  \Sigma_{ab} \left( x, y \right) = - \frac{i \lambda}{6 N} \delta_{ab} \delta_{\mathcal{C}} \left( x - y \right) G_{cc} \left( x, x \right) - \frac{i \lambda}{3 N} B^{-1} \left( x, y \right) G_{ab} \left( x, y \right) \;.
\end{equation}
By multiplying Eq.~(\ref{eq10}) from the left with $B^{-1}$ and using the identity
\begin{equation} \label{eq15}
  \intl_{\mathcal{C}} \dddd{z} \left[ B^{-1} \left( x, z \right) B \left( z, y \right) \right] = \delta_{\mathcal{C}} \left( x - y \right)
\end{equation}
we find for $B^{-1}$:
\begin{equation} \label{eq11}
  B^{-1} \left( x, y \right) = \delta_{\mathcal{C}} \left( x - y \right) - \frac{i \lambda}{6 N} \intl_{\mathcal{C}} \dddd{z} \left[ B^{-1} \left( x, z \right) G_{ab} \left( z, y \right) G_{ab} \left( z, y \right) \right] \;.
\end{equation}
From this we see, that we can write $B^{-1} (x, y)$ in the form
\begin{equation} \label{eq12}
  B^{-1} \left( x, y \right) = \delta_{\mathcal{C}} \left( x - y \right) - i I \left( x, y \right) \;,
\end{equation}
where $I (x, y)$ is given by Eq.~(\ref{eq111}). Further simplification arises 
if we take into account that for systems with $O(N)$-symmetric initial
conditions the propagator satisfies
\begin{equation} \label{eq85}
  G_{ab} \left( x, y \right) = G \left( x, y \right) \delta_{ab} \;.
\end{equation}
Eq.~(\ref{eq13}) shows that this property is inherited by the proper self 
energy, and  additionally it suggests, together with Eq.~(\ref{eq12}), to 
decompose the self energy into a local and a nonlocal part.
\[ \Sigma \left( x, y \right) = - i \Sigma^{\mbox{\scriptsize (local)}} \left( x \right) \delta_{\mathcal{C}} \left( x - y \right) + \Sigma^{\mbox{\scriptsize (nonlocal)}} \left( x, y \right) \]
The diagrammatic expansion of the proper self energy is illustrated in 
Fig.~\ref{fig4}. 
The two tadpoles in the self energy result from the LO and NLO double bubbles 
in $\Gamma_2$ and form the local part of the self energy. Correspondingly, 
the setting sun forms the nonlocal part. While the nonlocal part takes direct 
scattering into account, one immediately sees that the local part causes a 
mass shift only, wherefore we define the effective mass by
\begin{eqnarray}
        M^2 \left( x \right) 
  & = & m^2 + \Sigma^{(\mbox{\scriptsize local})} \left( x \right) \nonumber \\
  & = & m^2 + \lambda \frac{N + 2}{6 N} G \left( x, x \right) \label{eq74} \;.
\end{eqnarray}
When we replace $m^2$ in the Schwinger-Dyson equation (\ref{eq73}) by $M^2 
\left( x \right)$, the proper self energy reduces to its nonlocal part which 
results in
\begin{equation} \label{eq51}
  \Sigma^{(\mbox{\scriptsize nonlocal})} \left( x, y \right) = - \frac{\lambda}{3 N} G \left( x, y \right) I \left( x, y \right) \;,
\end{equation}
where due to Eqs.~(\ref{eq111}) and (\ref{eq85}) $I(x, y)$ satisfies
\begin{equation} \label{eq52}
  I \left( x, y \right) = \frac{\lambda}{6} G \left( x, y \right) G \left( x, y \right) - i \frac{\lambda}{6} \intl_{\mathcal{C}} \dddd{z} \left[ I \left( x, z \right) G \left( z, y \right) G \left( z, y \right) \right] \;.
\end{equation}
Next, we use the analogue to Eq.~(\ref{eq15}) for the propagator $G (x, y)$ 
to bring the Schwin\-ger-Dyson equation into a form that is appropriate for 
an initial value problem:
\begin{equation} \label{eq16}
  - i \delta_{\mathcal{C}} \left( x - y \right) = \left( \Box_x + M^2 \left( x \right) \right) G \left( x, y \right) + i \intl_{\mathcal{C}} \dddd{z} \left[ \Sigma^{(\mbox{\scriptsize nonlocal})} \left( x, z \right) G \left( z, y \right) \right] \;.
\end{equation}
Now, we decompose $G$ and $\Sigma^{(\mbox{\scriptsize nonlocal})}$ according
to Eq.~(\ref{eq1}), with $\Theta$ functions that are defined on the closed 
time path $\mathcal{C}$.
\begin{equation} \label{eq45}
  G \left( x, y \right) = \Theta_{\mathcal{C}} \left( x^0, y^0 \right) G_> \left( x, y \right) + \Theta_{\mathcal{C}} \left( y^0, x^0 \right) G_< \left( x, y \right)
\end{equation}
\begin{equation} \label{eq46}
  \Sigma^{(\mbox{\scriptsize nonlocal})} \left( x, y \right) = \Theta_{\mathcal{C}} \left( x^0, y^0 \right) \Sigma_> \left( x, y \right) + \Theta_{\mathcal{C}} \left( y^0, x^0 \right) \Sigma_< \left( x, y \right)
\end{equation}
Using that in the sense of distributions 
\[ \Box_x G \left( x, y \right) = \Theta_{\mathcal{C}} \left( x^0, y^0 \right) \Box_x G_> \left( x, y \right) + \Theta_{\mathcal{C}} \left( y^0, x^0 \right) \Box_x G_< \left( x, y \right) - i \delta_{\mathcal{C}} \left( x - y \right) \;, \]
we find that because of the decompositions (\ref{eq45}) and (\ref{eq46}), and 
the analy\-ticity properties of $G_> \left( x, y \right)$ and $G_< \left( x, y 
\right)$ with respect to their time arguments, Eq.~(\ref{eq16}) splits in 
two equations, one for $G_> \left( x, y \right)$ and one for $G_< \left( x, y 
\right)$\footnote{Here we also use the hermiticity of $G_> \left( x, y 
\right)$ and $G_< \left( x, y \right)$. The same relations hold for the 
proper self energy: $\coco{\Sigma}_> \left( x, y \right) = \Sigma_> \left( y, 
x \right) = \Sigma_< \left( x, y \right)$}:
\begin{eqnarray}
        \left( \Box_x + M^2 \left( x \right) \right) G_> \left( x, y \right)
  & = & 2 \int \ddd{z} \Bigg[ \intl_0^{x^0} \dd{z^0} \left[ \Im \left( \Sigma_> \left( x, z \right) \right) G_> \left( z, y \right) \right] \qquad \qquad \label{eq47} \\
  &   & {} - \intl_0^{y^0} \dd{z^0} \left[ \Sigma_> \left( x, z \right) \Im \left( G_> \left( z, y \right) \right) \right] \Bigg] \nonumber 
\end{eqnarray}
and
\begin{eqnarray}
        \left( \Box_x + M^2 \left( x \right) \right) G_< \left( x, y \right)
  & = & 2 \int \ddd{z} \Bigg[ \intl_0^{y^0} \dd{z^0} \left[ \Sigma_< \left( x, z \right) \Im \left( G_< \left( z, y \right) \right) \right] \qquad \qquad \label{eq48} \\
  &   & {} - \intl_0^{x^0} \dd{z^0} \left[ \Im \left( \Sigma_< \left( x, z \right) \right) G_< \left( z, y \right) \right] \Bigg] \nonumber \;.
\end{eqnarray}
It is worth noting that in the literature these equations are known as the 
Kada\-noff-Baym equations, and that they are used for the description of an 
enormous variety of different phenomena ranging from the very small as in our
case over conductors and semi-conductor devices to the very large as in
cosmology. Due to the hermiticity property of $G_> (x, y)$ and $G_< (x, y)$ 
one can reduce these two complex-valued evolution equations to two 
real-valued evolution equations. These two real-valued evolution equations 
are the evolution equations for the spectral function and the symmetric 
propagator which we mentioned earlier. Actually, the spectral function 
$\varrho (x, y)$ is a purely imaginary function. Therefore we define 
\[ \rho \left( x, y \right) \equiv i \varrho \left( x, y \right) \;, \]
which is obviously a real quantity. The evolution equation for the symmetric
propagator is given by the sum of Eqs.~(\ref{eq47}) and (\ref{eq48}), the one 
for the spectral function is obtained from their difference. 
\begin{myeqnarray}{E1}
        \left( \Box_x + M^2 \left( x \right) \right) F \left( x, y \right) 
  & = & - \int \ddd{z} \Bigg[ \intl_0^{x^0} \dd{z^0} \left[ \Sigma_{\rho} \left( x, z \right) F \left( z, y \right) \right] \label{eq50} \\
  &   & {} - \intl_0^{y^0} \dd{z^0} \left[ \Sigma_F \left( x, z \right) \rho \left( z, y \right) \right] \Bigg] \nonumber
\end{myeqnarray}
\begin{myequation}{E2} \label{eq49}
  \left( \Box_x + M^2 \left( x \right) \right) \rho \left( x, y \right) = \int \ddd{z} \intl_{x^0}^{y^0} \dd{z^0} \left[ \Sigma_{\rho} \left( x, z \right) \rho \left( z, y \right) \right]
\end{myequation}Because 
of Eq.~(\ref{eq51}) and the decomposition (\ref{eq46})\footnote{We 
also decompose the resummation function $I \left( x, y \right)$ according to\\
$I \left( x, y \right) = 
\Theta_{\mathcal{C}} \left( x^0, y^0 \right) I_> \left( x, y \right) + 
\Theta_{\mathcal{C}} \left( y^0, x^0 \right) I_< \left( x, y \right)$}
\begin{equation} \label{eq75}
  \Sigma_> \left( x, y \right) = - \frac{\lambda}{3 N} G_> \left( x, y \right) I_> \left( x, y \right)
\end{equation}
and
\begin{equation} \label{eq76}
  \Sigma_< \left( x, y \right) = - \frac{\lambda}{3 N} G_< \left( x, y \right) I_< \left( x, y \right) \;.
\end{equation}
For the resummation functions $I_> (x, y)$ and $I_< (x, y)$ we obtain
\begin{eqnarray*}
        I_> \left( x, y \right) 
  & = & \frac{\lambda}{6} G_> \left( x, y \right) G_> \left( x, y \right) \\
  &   & {} - \frac{i \lambda}{6} \int \ddd{z} \Bigg[ \intl_0^{x^0} \dd{z^0} \left[ \left( I_> \left( x, z \right) - I_< \left( x, z \right) \right) G_> \left( z, y \right) G_> \left( z, y \right) \right] \\
  &   & {} - \intl_0^{y^0} \dd{z^0} \left[ I_> \left( x, z \right) \left( G_> \left( z, y \right) G_> \left( z, y \right) - G_< \left( z, y \right) G_< \left( z, y \right) \right) \right] \Bigg]
\end{eqnarray*}
and
\begin{eqnarray*}
        I_< \left( x, y \right) 
  & = & \frac{\lambda}{6} G_< \left( x, y \right) G_< \left( x, y \right) \\
  &   & {} - \frac{i \lambda}{6} \int \ddd{z} \Bigg[ \intl_0^{x^0} \dd{z^0} \left[ \left( I_> \left( x, z \right) - I_< \left( x, z \right) \right) G_< \left( z, y \right) G_< \left( z, y \right) \right] \\
  &   & {} - \intl_0^{y^0} \dd{z^0} \left[ I_< \left( x, z \right) \left( G_> \left( z, y \right) G_> \left( z, y \right) - G_< \left( z, y \right) G_< \left( z, y \right) \right) \right] \Bigg] \;.
\end{eqnarray*}
The various self energies that appear in the evolution equations (\ref{eq50})
and (\ref{eq49}) are then given by
\begin{myeqnarray}{E3}
        \Sigma_F \left( x, y \right) 
  & = & \frac{1}{2} \left( \Sigma_> \left( x, y \right) + \Sigma_< \left( x, y \right) \right) \nonumber \\
  & = & - \frac{\lambda}{3 N} \left( F \left( x, y \right) I_F \left( x, y \right) - \frac{1}{4} \rho \left( x, y \right) I_{\rho} \left( x, y \right) \right) \label{eq17}
\end{myeqnarray}and
\begin{myeqnarray}{E4}
        \Sigma_{\rho} \left( x, y \right)
  & = & i \left( \Sigma_> \left( x, y \right) - \Sigma_< \left( x, y \right) \right) \nonumber \\
  & = & - \frac{\lambda}{3 N} \Big( F \left( x, y \right) I_{\rho} \left( x, y \right) + \rho \left( x, y \right) I_{F} \left( x, y \right) \Big) \;, \label{eq18} 
\end{myeqnarray}where 
the symmetric and the spectral resummation functions $I_F (x, y)$ and 
$I_{\rho} (x, y)$ are defined by
\begin{myeqnarray}{E5}
        I_F \left( x, y \right)
  & = & \frac{1}{2} \left( I_> \left( x, y \right) + I_< \left( x, y \right) \right) \nonumber \\
  & = & \frac{\lambda}{6} \left( F^2 \left( x, y \right) - \frac{1}{4} \rho^2 \left( x, y \right) \right) \nonumber \\
  &   & {} - \frac{\lambda}{6} \int \ddd{z} \Bigg[ \intl_0^{x^0} \dd{z^0} \bigg[ I_{\rho} \left( x, z \right) \bigg( F^2 \left( z, y \right) - \frac{1}{4} \rho^2 \left( z, y \right) \bigg) \bigg] \nonumber \\
  &   & {} - \intl_0^{y^0} \dd{z^0} \left[ I_F \left( x, z \right) F \left( z, y \right) \rho \left( z, y \right) \right] \Bigg] \label{eq86}
\end{myeqnarray}and
\begin{myeqnarray}{E6}
        I_{\rho} \left( x, y \right) 
 & = & i \left( I_> \left( x, y \right) - I_< \left( x, y \right) \right) \nonumber \\
 & = & \frac{\lambda}{3} F \left( x, y \right) \rho \left( x, y \right) \label{eq87} \\
 &   & {} - \frac{\lambda}{3} \int \ddd{z} \intl_{y^0}^{x^0} \dd{z^0} \left[ I_{\rho} \left( x, z \right) F \left( z, y \right) \rho \left( z, y \right) \right] \;. \nonumber
\end{myeqnarray}The 
evolution equations (\ref{eq50}) to (\ref{eq87}) form a closed set of 
equations which determine the evolution of the Schwinger-Keldysh propagator
completely. It is worth noting that these equations are explicit in time. 
This is obvious for the self energies. For those equations containing 
integrals it means that, when evaluating such an integral for given $x^0$ and
$y^0$, one only needs to know all the functions in the integrand for times 
between the initial time and $x^0$ or $y^0$, respectively. This allows for an
efficient numerical solution of these equations for any given initial 
conditions. A detailed discussion of this as well as numerically obtained 
solutions for miscellaneous initial conditions can be found in 
Ref.~\cite{Be1}.

We mentioned in the introduction that the $1/N$-expansion of the 2PI 
effective action is capable of controlling nonperturbatively large 
fluctuations. Now, we are able to prove this statement \cite{BeSe,BeMu}. In 
the case of nonperturbatively large fluctuations
\[ F \left( x, y \right) \sim \mathcal{O} \left( \frac{1}{\lambda} \right) \qquad \mbox{and} \qquad \rho \left( x, y \right) \sim \mathcal{O} \left( 1 \right) \;. \]
One sees immediately, that this property is inherited by the resummation 
functions $I_F \left( x, y \right)$ and $I_{\rho} \left( x, y \right)$. As a 
consequence every diagram contributing to the $1/N$-expansion of the 2PI
effective action to next-to-leading order in the symmetric regime 
contributes with the same order in the coupling. Therefore any loop expansion 
breaks down in such a situation and one has to use a nonperturbative approach, 
as provided by the $1/N$-expansion of the 2PI effective action.

In the next chapter, using a combination of a first order gradient expansion 
and a Wigner transformation, we are going to derive kinetic equations from 
these evolution equations. In particular, by additionally employing a 
quasi-particle (on-shell) approximation, we are going to show how one can
obtain a generalized Boltzmann equation. In this sense one can consider the
full evolution equations as {\em quantum Boltzmann equations} resumming the
gradient expansion up to infinite order and including off-shell effects 
\cite{AaBe1}.

\chapter{Local Approach to Equilibrium}
\thispagestyle{empty}

In this chapter we consider quantum fields in the intermediate-time drifting 
and the late-time thermalization regimes. As pointed out in the 
introduction we will exploit the slight dependence of the
propagator on the center coordinates and the effective memory loss. By 
combining a first order gradient expansion and a Wigner transformation we 
derive kinetic equations that are local in time from the full nonlocal
evolution equations. These kinetic equations are important tools for the 
description of the evolution of nonequilibrium quantum fields for large times 
where the memory integrals complicate computations. 

Additionally employing a quasi-particle approximation, we obtain a 
generalized Boltzmann equation. In the derivation of the classical Boltzmann
equation one assumes that the system is so dilute such that there are 
practically only 
collisions which involve two particles and that one can neglect collisions 
between three or more particles. In contrast to the classical Boltzmann
equation our generalized Boltzmann equation includes an effective mass and an
effective coupling which arise from the resummation of the infinite series of 
diagrams that contribute to the $1/N$-expansion of the 2PI effective action 
to next-to-leading order. Due to the effective coupling we expect our 
generalized Boltzmann equation to be valid even for systems which are so dense
that one cannot neglect collisions between three or more particles.

\section{Quantum Kinetic Equations}

We begin the derivation of the kinetic equations with the evolution equations 
for the symmetric propagator and the spectral function (\ref{eq50}) and 
(\ref{eq49}), namely
\begin{eqnarray}
        \lefteqn{\left( \Box_x + M^2 \left( x \right) \right) F \left( x, y \right)} \; \label{eq37} \\
  & = & - \int \dddd{z} \bigg[ \theta \left( z^0 \right) \Big( \Sigma_R \left( x , z \right) F \left( z, y \right) + \Sigma_F \left( x, z \right) G_A \left( z, y \right) \Big) \bigg] \nonumber
\end{eqnarray}
and
\begin{eqnarray}
        \lefteqn{\left( \Box_x + M^2 \left( x \right) \right) \varrho \left( x, y \right)} \; \label{eq38} \\
  & = & - \int \dddd{z} \bigg[ \theta \left( z^0 \right) \Big( \Sigma_{\varrho} \left( x , z \right) G_A \left( z, y \right) + \Sigma_R \left( x, z \right) \varrho \left( z, y \right) \Big) \bigg] \nonumber \;.
\end{eqnarray}
Now, we interchange $x$ and $y$ in Eq.~(\ref{eq37}) and use the symmetry 
properties of the symmetric propagator and the self energies with respect to 
the order of their arguments. Subtracting the obtained equation from 
Eq.~(\ref{eq37}), we get:
\begin{eqnarray}
        \lefteqn{\Big( \Box_x - \Box_y + M^2 \left( x \right) - M^2 \left( y \right) \Big) F \left( x, y \right)} \; \label{eq40} \\
  & = & \int \dddd{z} \bigg[ \theta \left( z^0 \right) \Big( F \left( x, z \right) \Sigma_A \left( z, y \right) + G_R \left( x, z \right) \Sigma_F \left( z, y \right) \nonumber \\ 
  &   & {} - \Sigma_R \left( x , z \right) F \left( z, y \right) - \Sigma_F \left( x, z \right) G_A \left( z, y \right) \Big) \bigg] \nonumber \;.
\end{eqnarray}
The same procedure yields for Eq.~(\ref{eq38}):
\begin{eqnarray}
        \lefteqn{\Big( \Box_x - \Box_y + M^2 \left( x \right) - M^2 \left( y \right) \Big) \varrho \left( x, y \right)} \; \label{eq88} \\
  & = & \int \dddd{z} \bigg[ \theta \left( z^0 \right) \Big( G_R \left( x, z \right) \Sigma_{\varrho} \left( z, y \right) + \varrho \left( x, z \right) \Sigma_A \left( z, y \right) \nonumber \\ 
  &   & {} - \Sigma_{\varrho} \left( x , z \right) G_A \left( z, y \right) - \Sigma_R \left( x, z \right) \varrho \left( z, y \right) \Big) \bigg] \nonumber \;.
\end{eqnarray}
For every two points $v$ and $w$ in space time, their center and relative
coordinates are denoted by
\[ X_{vw} = \frac{v + w}{2} \qquad \mbox{and} \qquad s_{vw} = v - w \;. \]
However, the space time coordinates $x$ and $y$ on the left hand sides of 
Eqs.~(\ref{eq40}) and (\ref{eq88}) are particularly important:
\begin{equation} \label{eq71}
  X \equiv X_{xy} = \frac{x + y}{2} \qquad \mbox{and} \qquad s \equiv s_{xy} = x - y \;.
\end{equation}
Every function $f$ of $v$ and $w$ can be written as a function of the center
coordinate $X_{vw}$ and the relative coordinate $s_{vw}$:
\[ f \left( v, w \right) = f \left( X_{vw} + \frac{s_{vw}}{2}, X_{vw} - \frac{s_{vw}}{2} \right) = \tilde{f} \left( X_{vw}, s_{vw} \right) \;. \]
We now rewrite Eqs.~(\ref{eq40}) and (\ref{eq88}) using center and relative 
coordinates. An application of the chain rule gives
\[ \Box_x - \Box_y = \partial_{x^{\mu}} \partial_{x_{\mu}} - \partial_{y^{\mu}} \partial_{y_{\mu}} = 2 \partial_{s_{\mu}} \partial_{X^{\mu}} \;. \]
In the drifting regime, the exponential suppression of correlations with 
the initial time allows us to send the initial time to $- \infty$, i.e. from 
here on we drop the $\theta$ function in the above memory integrals. Hence, 
the expressions on the right hand sides of Eqs.~(\ref{eq40}) and (\ref{eq88})
include integrals of the form
\begin{equation} \label{eq4}
  \int \dddd{z} \Big[ f \left( x, z \right) g \left( z, y \right) \Big] = \int \dddd{z} \Big[ \tilde{f} \left( X_{xz}, s_{xz} \right) \tilde{g} \left( X_{zy}, s_{zy} \right) \Big] \;.
\end{equation}
In equilibrium the propagator and the proper self energy are invariant under
space-time translations. This means that they depend on the relative
coordinates only, and that there is no dependence on the center coordinates. 
This does not hold for a system out of thermal equilibrium, where the 
propagator and the self energy depend on the center coordinates. However, if 
the propagator is a sufficiently smooth function of the center coordinates
--- which it is by definition in the intermediate-time drifting regime 
(see Fig.~\ref{fig7}) --- a Taylor expansion to first order in the center 
coordinates should be a good approximation. At this point we would
like to emphasize that the system still may be far away from thermal 
equilibrium and that we do not justify the Taylor expansion by assuming that
the system has approached the equilibrium sufficiently closely. The Taylor 
expansion of the mass terms to first order in $\partial_{X^{\mu}}$ around $X$ 
gives
\[ M^2 \left( X + \frac{s}{2} \right) - M^2 \left( X - \frac{s}{2} \right) = s^{\mu} \cdot \left( \partial_{X^{\mu}} M^2 \left( X \right) \right) \;. \]
In the integral (\ref{eq4}) the expansion of  $\tilde{f}$ and $\tilde{g}$  
into a Taylor series to first order in $\partial_{X^{\mu}}$ around 
$X_{xz} = X$ and $X_{zy} = X$ respectively, yields
\begin{eqnarray*}
        \lefteqn{\int \dddd{z} \Big[ f \left( x, z \right) g \left( z, y \right) \Big]} \; \\
  & = & \int \dddd{z} \Big[ \tilde{f} \left( X, s_{xz} \right) \tilde{g} \left( X, s_{zy} \right) - \frac{s_{xz}^{\mu}}{2} \tilde{f} \left( X, s_{xz} \right) \left( \partial_{X^{\mu}} \tilde{g} \left( X, s_{zy} \right) \right) \\
  &   & {} + \frac{s_{zy}^{\mu}}{2} \left( \partial_{X^{\mu}} \tilde{f} \left( X, s_{xz} \right) \right) \tilde{g} \left( X, s_{zy} \right) \Big] \;.
\end{eqnarray*}
After we have removed the $\theta$ functions from the memory integrals and 
performed the 
first order gradient expansion in Eqs.~(\ref{eq40}) and (\ref{eq88}), our next 
step is to Fourier transform these equations with respect to $s$. On the left 
hand side we get
\begin{eqnarray*}
        \lefteqn{LHS \left( \ref{eq40} \right)} \; \\
  & \to & \int \dddd{s} \Bigg[ \exp \left( i k s \right) \left( 2 \partial_{s_{\mu}} \partial_{X^{\mu}} + s^{\mu} \left( \partial_{X^{\mu}} M^2 \left( X \right) \right) \right) \tilde{F} \left( X, s \right) \Bigg] \\
  & = & - i \left( 2 k^{\mu} \partial_{X^{\mu}} + \left( \partial_{X^{\mu}} M^2 \left( X \right) \right) \partial_{k_{\mu}} \right) \tilde{F} \left( X, k \right)
\end{eqnarray*}
and similar for Eq.~(\ref{eq88}). $\tilde{F} \left( X, k \right)$ denotes the 
so-called {\em Wigner transform} of $F \left( x, y \right)$, which is 
nothing but the Fourier transform of $\tilde{F} \left( X, s \right)$ with 
respect to $s$. It is worth noting that due to the hermiticity of $G_> \left( 
x, y \right)$ and $G_< \left( x, y \right)$, the Wigner transforms of the 
symmetric propagator and the spectral function are real functions. 

On the right hand side of Eqs.~(\ref{eq40}) and (\ref{eq88}), we find two 
different types of integrals, which we transform individually:
\begin{eqnarray*}
        \lefteqn{\int \dddd{z} \left[ \tilde{f} \left( X, s_{xz} \right) \tilde{g} \left( X, s_{zy} \right) \right]} \; \\
  & \to & \int \dddd{s} \int \dddd{z} \Big[ \exp \left( iks \right) \tilde{f} \left( X, s_{xz} \right) \tilde{g} \left( X, s_{zy} \right) \Big] \\
  & = & \tilde{f} \left( X, k \right) \tilde{g} \left( X, k \right)
\end{eqnarray*}
and
\begin{eqnarray*}
        \lefteqn{\int \dddd{z} \left[ - \frac{s_{xz}^{\mu}}{2} \tilde{f} \left( X, s_{xz} \right) \left( \partial_{X^{\mu}} \tilde{g} \left( X, s_{zy} \right) \right) + \frac{s_{zy}^{\mu}}{2} \left( \partial_{X^{\mu}} \tilde{f} \left( X, s_{xz} \right) \right) \tilde{g} \left( X, s_{zy} \right) \right]} \; \\
  & \to & \int \dddd{s} \int \dddd{z} \left[ \exp \left( iks \right) \left( - \frac{s_{xz}^{\mu}}{2} \tilde{f} \left( \partial_{X^{\mu}} \tilde{g} \right) + \frac{s_{zy}^{\mu}}{2} \left( \partial_{X^{\mu}} \tilde{f} \right) \tilde{g} \right) \right] \\
  & = & \frac{i}{2} \left( \left( \partial_{k_{\mu}} \tilde{f} \left( X, k \right) \right) \left( \partial_{X^{\mu}} \tilde{g} \left( X, k \right) \right) - \left( \partial_{X^{\mu}} \tilde{f} \left( X, k \right) \right) \left( \partial_{k_{\mu}} \tilde{g} \left( X, k \right) \right) \right) \\
  & \equiv & \frac{i}{2} \left\{ \tilde{f} \left( X, k \right) ; \tilde{g} \left( X, k \right) \right\}_{PB} \;.
\end{eqnarray*}
In the last step, we have introduced the Poisson brackets to shorten the 
notation. After all these transformations Eqs.~(\ref{eq40}) and (\ref{eq88})
become
\begin{eqnarray} 
        \lefteqn{- i \left( 2 k^{\mu} \partial_{X^{\mu}} + \left( \partial_{X^{\mu}} M^2 \left( X \right) \right) \partial_{k_{\mu}} \right) \tilde{F} \left( X, k \right)} \; \label{eq19} \\
  & = & - \tilde{F} \left( \tilde{\Sigma}_R - \tilde{\Sigma}_A \right) + \tilde{\Sigma}_F \left( \tilde{G}_R - \tilde{G}_A \right) \nonumber \\
  &   & {} + \frac{i}{2} \left\{ \tilde{F} ; \tilde{\Sigma}_R + \tilde{\Sigma}_A \right\}_{PB} - \frac{i}{2} \left\{ \tilde{\Sigma}_F ; \tilde{G}_R + \tilde{G}_A \right\}_{PB} \nonumber
\end{eqnarray}
and
\begin{eqnarray} 
        \lefteqn{- i \left( 2 k^{\mu} \partial_{X^{\mu}} + \left( \partial_{X^{\mu}} M^2 \left( X \right) \right) \partial_{k_{\mu}} \right) \tilde{\varrho} \left( X, k \right)} \; \label{eq41} \\
  & = & \tilde{\Sigma}_{\varrho} \left( \tilde{G}_R - \tilde{G}_A \right) - \tilde{\varrho} \left( \tilde{\Sigma}_R - \tilde{\Sigma}_A \right) \nonumber \\
  &   & {} - \frac{i}{2} \left\{ \tilde{\Sigma}_{\varrho} ; \tilde{G}_R + \tilde{G}_A \right\}_{PB} + \frac{i}{2} \left\{ \tilde{\varrho} ; \tilde{\Sigma}_R + \tilde{\Sigma}_A \right\}_{PB} \nonumber \;,
\end{eqnarray}
where all the functions on the right hand side depend on $X$ and $k$. From 
Fourier's theory, we know that for two functions $\tilde{f} \left( X, s 
\right)$ and $\tilde{g} \left( X, s \right)$ the Fourier transform of their 
product with respect to $s$ is given by the convolution of their individual
Fourier transforms
\begin{eqnarray*}
        \int \dddd{s} \left[ \exp \left( iks \right) \tilde{f} \left( X, s \right) \tilde{g} \left( X, s \right) \right] 
  & = & \int \frac{\dddd{p}}{\left( 2 \pi \right)^4} \left[ \tilde{f} \left( X, k - p \right) \tilde{g} \left( X, p \right) \right] \\
  & \equiv & \left( \tilde{f} \ast \tilde{g} \right) \left( X, k \right) \;,
\end{eqnarray*}
and that the Fourier transform of a $\theta$ function in one dimension is given 
by
\[ \int \dd{x} \left[ \exp \left( ikx \right) \theta \left( x \right) \right] = \lim_{\epsilon \to 0} \frac{i}{k + i \epsilon} \;. \]
Hence, we can write the Wigner transforms of the retarded and the advanced 
propagator in the following form:
\begin{equation} \label{eq25}
  \tilde{G}_R \left( X, k \right) = - \int \frac{\dd{p^0}}{2 \pi} \left[ \frac{\tilde{\varrho} \left( X, \bm{k}, p^0 \right)}{k^0 - p^0 + i \epsilon} \right] \;,
\end{equation}
\begin{equation} \label{eq39}
  \tilde{G}_A \left( X, k \right) = - \int \frac{\dd{p^0}}{2 \pi} \left[ \frac{\tilde{\varrho} \left( X, \bm{k}, p^0 \right)}{k^0 - p^0 - i \epsilon} \right] \;.
\end{equation}
As the $\delta$ function can be approximated by
\[ \delta_{\epsilon} \left( k^0 \right) = \frac{\epsilon}{\pi \left( k_0^2 + \epsilon^2 \right)} \;, \]
we find that
\begin{equation} \label{eq34}
 \tilde{G}_R \left( X, k \right) - \tilde{G}_A \left( X, k \right) = i \tilde{\varrho} \left( X, k \right) \;.
\end{equation}
The same relation holds for the proper self energy. On the other hand, we have 
already seen that $\tilde{\varrho} \left( X, k \right)$ is a real function. 
Therefore, Eqs.~(\ref{eq25}) and (\ref{eq39}) yield
\[ \coco{\tilde{G}_R} \left( X, k \right) = \tilde{G}_A \left( X, k \right) \;. \]
From this we see immediately that 
\begin{equation} \label{eq83}
  \tilde{G}_R \left( X, k \right) + \tilde{G}_A \left( X, k \right) = 2 \cdot \Re \left( \tilde{G}_R \left( X, k \right) \right)
\end{equation}
and
\begin{equation} \label{eq84}
  \tilde{\varrho} \left( X, k \right) = 2 \cdot \Im \left( \tilde{G}_R \left( X, k \right) \right) \;.
\end{equation}
Again, the same relations hold for the self energy, and Eqs.~(\ref{eq19}) and
(\ref{eq41}) now have been simplified to take the following form:
\begin{myeqnarray}{K1}
        \lefteqn{\left( 2 k^{\mu} \partial_{X^{\mu}} + \left( \partial_{X^{\mu}} M^2 \left( X \right) \right) \partial_{k_{\mu}} \right) \tilde{F} \left( X, k \right)} \; \nonumber \\
  & = & \tilde{F} \left( X, k \right) \tilde{\Sigma}_{\varrho} \left( X, k \right) - \tilde{\varrho} \left( X, k \right) \tilde{\Sigma}_F \left( X, k \right) \label{eq31} \\
  &   & {} + \left\{ \tilde{\Sigma}_F \left( X, k \right) , \Re \left( \tilde{G}_R \left( X, k \right) \right) \right\}_{PB} + \left\{ \Re \left( \tilde{\Sigma}_R \left( X, k \right) \right) , \tilde{F} \left( X, k \right) \right\}_{PB} \nonumber
\end{myeqnarray}and
\begin{myeqnarray}{K2}
        \lefteqn{\left( 2 k^{\mu} \partial_{X^{\mu}} + \left( \partial_{X^{\mu}} M^2 \left( X \right) \right) \partial_{k_{\mu}} \right) \tilde{\varrho} \left( X, k \right)} \; \label{eq30} \\
  & = & \left\{ \tilde{\Sigma}_{\varrho} \left( X, k \right) , \Re \left( \tilde{G}_R \left( X, k \right) \right) \right\}_{PB} + \left\{ \Re \left( \tilde{\Sigma}_R \left( X, k \right) \right) , \tilde{\varrho} \left( X, k \right) \right\}_{PB} \nonumber \;.
\end{myeqnarray}These 
are the kinetic equations for the symmetric propagator and the spectral 
function which we mentioned earlier.
In the next section we will show that, under certain assumptions, one can 
derive a generalized Boltzmann equation from Eq.~(\ref{eq31}). As we will see, 
the left hand side of Eq.~(\ref{eq31}) will reduce to the usual form of the 
left hand side of the classical Boltzmann equation as for example presented 
in the third section of Ref.~\cite{La10}, and on the right hand side 
$\tilde{F} \tilde{\Sigma}_{\varrho} - \tilde{\varrho} \tilde{\Sigma}_F$ will 
evoke a collision integral.

The convolution of the Wigner transformed spectral function with the Wigner 
transformed $\theta$ function as seen in Eq.~(\ref{eq25}) leads to a principal 
value integral which in general cannot be evaluated. Therefore, we cannot use 
Eq.~(\ref{eq25}) to calculate the Wigner transform of the retarded propagator, 
which appears in the kinetic equations (\ref{eq31}) and (\ref{eq30}). We 
circumvent this difficulty by deriving an extra equation for the Wigner 
transform of the retarded propagator. For the same reason we will derive a 
similar equation for the Wigner transformed retarded self energy as well. We 
obtain the equation for the Wigner transform of the retarded propagator in 
the following way: By calculating the effect of $\Box_x$ on $G_R \left( x, y 
\right)$ and $G_A \left( x, y \right)$ we find
\[ i \theta \left( x^0 - y^0 \right) \Box_x \varrho \left( x, y \right) = \Box_x G_R \left( x, y \right) - \delta \left( x - y \right) \;, \]
\[ - i \theta \left( y^0 - x^0 \right) \Box_x \varrho \left( x, y \right) = \Box_x G_A \left( x, y \right) - \delta \left( y - x \right) \;. \]
From this we see that the multiplication of Eq.~(\ref{eq38}) once with 
$i \theta \left( x^0 - y^0 \right)$ and once with $- i \theta \left( y^0 - 
x^0 \right)$ yields equations for $G_R \left( x, y \right)$ and $G_A \left( 
x, y \right)$ respectively:
\begin{equation} \label{eq21}
  \delta \left( x - y \right) = \left( \Box_x + M^2 \left( x \right) \right) G_R \left( x, y \right) + \int \dddd{z} \left[ \Sigma_R \left( x, z \right) G_R \left( z, y \right) \right]
\end{equation}
and
\begin{equation} \label{eq22}
  \delta \left( y - x \right) = \left( \Box_x + M^2 \left( x \right) \right) G_A \left( x, y \right) + \int \dddd{z} \left[ \Sigma_A \left( x, z \right) G_A \left( z, y \right) \right] \;.
\end{equation}
We add Eq.~(\ref{eq22}), with $x$ and $y$ interchanged, to Eq.~(\ref{eq21}), 
which gives
\begin{eqnarray}
  2 \delta \left( x - y \right) & = & \left( \Box_x + \Box_y + M^2 \left( x \right) + M^2 \left( y \right) \right) G_R \left( x, y \right) \label{eq23} \\
                                &   & {} + \int \dddd{z} \left[ \Sigma_R \left( x, z \right) G_R \left( z, y \right) +  G_R \left( x, z \right) \Sigma_R \left( z, y \right) \right] \nonumber \;,
\end{eqnarray}
and proceed along the lines that lead from Eqs.~(\ref{eq40}) and (\ref{eq88}) 
to the kinetic equations (\ref{eq31})
and (\ref{eq30}). When, we rewrite Eq.~(\ref{eq23}) using center and relative 
coordinates and perform a first order gradient expansion, we find that the 
effective masses give
\[ M^2 \left( x \right) + M^2 \left( y \right) = 2 M^2 \left( X \right) \;. \]
Furthermore, neglecting terms beyond first order in $\partial_{X^{\mu}}$, we
obtain
\[ \Box_x + \Box_y = 2 \partial_{s^{\mu}} \partial_{s_{\mu}} \;. \]
The integral on the right hand side behaves exactly in the same way as the
ones that we met during the derivation of the kinetic equations (\ref{eq31})
and (\ref{eq30}). Finally we perform a Fourier transformation with respect to 
$s$. The result is an algebraic equation for $\tilde{G}_R \left( X, k \right)$:
\begin{myequation}{K3} \label{eq24} 
  1 = \left( - k^2 + M^2 \left( X \right) + \tilde{\Sigma}_R \left( X, k \right) \right) \tilde{G}_R \left( X, k \right) \;.
\end{myequation}Due 
to Eqs.~(\ref{eq24}) and (\ref{eq34}), we can express the spectral 
function in terms of the proper self energy:
\begin{equation} \label{eq35}
  \tilde{\varrho} \left( X, k \right) = \frac{- \tilde{\Sigma}_{\varrho} \left( X, k \right)}{\left( - k^2 + M^2 \left( X \right) + \Re \left( \tilde{\Sigma}_R \left( X, k \right) \right) \right)^2 + \left( \frac{1}{2} \tilde{\Sigma}_{\varrho} \left( X, k \right) \right) ^2} \;.
\end{equation}
We still need to work out the expressions of the self energies that appear in 
the kinetic equations. In addition to the expressions (\ref{eq17}) and 
(\ref{eq18}) for the symmetric and the spectral self energy we find for the 
retarded self energy:
\begin{eqnarray}
          \Sigma_R \left( x, y \right)
  & = & i \theta \left( x^0 - y^0 \right) \Sigma_{\varrho} \left( x, y \right) \nonumber \\
  & = & - \frac{\lambda}{3 N} \Big( F \left( x, y \right) I_R \left( x, y \right) + G_R \left( x, y \right) I_{F} \left( x, y \right) \Big) \label{eq89} \;.
\end{eqnarray}
The Wigner transformation of Eqs.~(\ref{eq17}), (\ref{eq18}) and (\ref{eq89})
yields:
\begin{myequation}{K4}
  \tilde{\Sigma}_F \left( X, k \right) = - \frac{\lambda}{3 N} \left( \left( \tilde{F} \ast \tilde{I}_F \right) \left( X, k \right) + \frac{1}{4} \left( \tilde{\varrho} \ast \tilde{I}_{\varrho} \right) \left( X, k \right) \right) \;,
\end{myequation}
\begin{myequation}{K5}
  \tilde{\Sigma}_{\varrho} \left( X, k \right) = - \frac{\lambda}{3 N} \bigg( \left( \tilde{F} \ast \tilde{I}_{\varrho} \right) \left( X, k \right) + \left( \tilde{\varrho} \ast \tilde{I}_F \right) \left( X, k \right) \bigg) \;,
\end{myequation}
\begin{myequation}{K6} \label{eq106}
  \tilde{\Sigma}_R \left( X, k \right) = - \frac{\lambda}{3 N} \Big( \left( \tilde{F} \ast \tilde{I}_R \right) \left( X, k \right) + \left( \tilde{G}_R \ast \tilde{I}_F \right) \left( X, k \right) \Big) \;.
\end{myequation}In 
the derivation of the kinetic equations we exploited the effective memory
loss in order to drop the $\theta$ functions that appear in the full evolution 
equations (\ref{eq37}) and (\ref{eq38}). By the same reasoning we can drop 
the respective $\theta$ functions in the equations for $I_F (x, y)$ and 
$I_{\varrho} (x, y)$ which then read:
\begin{eqnarray}
        I_F \left( x, y \right)
  & = & \frac{\lambda}{6} \left( F^2 \left( x, y \right) + \frac{1}{4} \varrho^2 \left( x, y \right) \right) \nonumber \\
  &   & {} - \frac{\lambda}{6} \int \dddd{z} \Bigg[ I_R \left( x, z \right) \bigg( F^2 \left( z, y \right) + \frac{1}{4} \varrho^2 \left( z, y \right) \bigg) \nonumber \\
  &   & {} + 2 I_F \left( x, z \right) F \left( z, y \right) G_A \left( z, y \right) \Bigg] \label{eq26}
\end{eqnarray}
and
\begin{eqnarray}
        I_{\varrho} \left( x, y \right)
  & = & \frac{\lambda}{3} F \left( x, y \right) \varrho \left( x, y \right) \nonumber \\
  &   & {} - \frac{\lambda}{3} \int \dddd{z} \Bigg[ I_{\varrho} \left( x, z \right) F \left( z, y \right) G_A \left( z, y \right) \nonumber \\
  &   & {} + I_R \left( x, z \right) F \left( z, y \right) \varrho \left( z, y \right) \Bigg] \label{eq27} \;.
\end{eqnarray}
Multiplying Eq.~(\ref{eq27}) with $i \theta \left( x^0 - y^0 \right)$ gives an 
equation for the retarded resummation function $I_R \left( x, y \right)$:
\begin{eqnarray}
        I_R \left( x, y \right) 
  & = & \frac{\lambda}{3} G_R \left( x, y \right) F \left( x, y \right) \label{eq77} \\
  &   & {} - \frac{\lambda}{3} \int \dddd{z} \left[ I_R \left( x, z \right) G_R \left( z, y \right) F \left( z, y \right) \right] \nonumber \;.
\end{eqnarray}
When we rewrite Eqs.~(\ref{eq26}), (\ref{eq27}) and (\ref{eq77}) using center 
and relative coordinates, expand the functions inside the integrals to first 
order in $\partial_{X^{\mu}}$ around $X$ and Fourier transform the equations 
with respect to $s$, we obtain the following equations for the Wigner 
transformed resummation functions:
\begin{myeqnarray}{K7}
        \tilde{I}_F \left( X, k \right) 
  & = & \frac{\lambda}{6} \left( \left( \tilde{F} \ast \tilde{F} \right) + \frac{1}{4} \left( \tilde{\varrho} \ast \tilde{\varrho} \right) \right) \label{eq59} \\
  &   & {} - \frac{\lambda}{6} \Bigg[ \tilde{I}_R \cdot \left( \left( \tilde{F} \ast \tilde{F} \right) + \frac{1}{4} \left( \tilde{\varrho} \ast \tilde{\varrho} \right) \right) + 2 \tilde{I}_F \left( \tilde{F} \ast \tilde{G}_A \right) \nonumber \\
  &   & {} + \frac{i}{2} \left\{ \tilde{I}_R ; \left( \tilde{F} \ast \tilde{F} \right) + \frac{1}{4} \left( \tilde{\varrho} \ast \tilde{\varrho} \right) \right\}_{PB} + i \left\{ \tilde{I}_F ; \left( \tilde{F} \ast \tilde{G}_A \right) \right\}_{PB} \Bigg] \nonumber \;,
\end{myeqnarray}
\begin{myeqnarray}{K8}
        \tilde{I}_{\varrho} \left( X, k \right) 
  & = & \frac{\lambda}{3} \left( \tilde{F} \ast \tilde{\varrho} \right) - \frac{\lambda}{3} \Bigg[ \tilde{I}_R \left( \tilde{F} \ast \tilde{\varrho} \right) + \tilde{I}_{\varrho} \left( \tilde{F} \ast \tilde{G}_A \right) \label{eq60} \\
  &   & + \frac{i}{2} \left\{ \tilde{I}_R ; \left( \tilde{F} \ast \tilde{\varrho} \right) \right\}_{PB} + \frac{i}{2} \left\{ \tilde{I}_{\varrho} ; \left( \tilde{F} \ast \tilde{G}_A \right) \right\}_{PB} \Bigg] \nonumber \;,
\end{myeqnarray}
\begin{myequation}{K9} \label{eq61}
  \tilde{I}_R \left( X, k \right) = \frac{\lambda}{3} \left( \tilde{G}_R \ast \tilde{F} \right) \left( 1 - \tilde{I}_R \right) + \frac{i \lambda}{6} \left\{ \left( \tilde{G}_R \ast \tilde{F} \right) ; \tilde{I}_R \right\}_{PB} \;.
\end{myequation}Of 
course, the functions and convolutions on the right hand sides
depend on the center coordinate $X$ and the momentum $k$, without exception. 

We stress that the kinetic equations (\ref{eq31}) to (\ref{eq61})
form --- exactly as the evolution equations (\ref{eq50}) to (\ref{eq87}) in 
Sect.~\ref{sect2} --- a closed set of equations. Their range of applicability 
is restricted only by the gradient expansion that we employed in their 
derivation. Hence, in regimes where this gradient expansion is justified, the 
kinetic equations (\ref{eq31}) to (\ref{eq61}) describe the evolution of 
nonequilibrium
quantum fields without further approximations. In this context we would like 
to emphasize that our kinetic equations indeed inherit the capability of 
describing nonperturbatively large fluctuations from the nonlocal evolution 
equations, as it was described in Sect.~\ref{sect2} \cite{BeMu}. Furthermore, 
it is important to note that we derived our kinetic equations starting from 
the effective action. As a consequence it can be shown that our kinetic 
equations possess the generic feature of exact conservation laws at the 
level of expectation values \cite{IvKnVo3}.

Before we close this section, we convince ourselves that our kinetic 
equations contain a consistent description of the equilibrium. We mentioned 
already in Sect.~\ref{sect1} that in equilibrium the propagator, and thus the 
proper self energy too, are space-time translation invariant. Hence, they 
do not depend on the center coordinate $X$, and so the left hand sides of 
Eqs.~(\ref{eq31}) and (\ref{eq30}) as well as the Poisson brackets on their 
right hand sides vanish. From this we see already that at least 
Eq.~(\ref{eq30}) is identically fulfilled in equilibrium. To see that
Eq.~(\ref{eq31}) also describes the equilibrium in a consistent way, it 
remains to show that the collision term also vanishes in equilibrium. To see 
this most easily, we note that we can write the collision term also as
\begin{equation} \label{eq20}
  \tilde{F} \left( k \right) \tilde{\Sigma}_{\varrho} \left( k \right) - \tilde{\varrho} \left( k \right) \tilde{\Sigma}_F \left( k \right) = \tilde{G}_< \left( k \right) \tilde{\Sigma}_> \left( k \right) - \tilde{G}_> \left( k \right) \tilde{\Sigma}_< \left( k \right) \;.
\end{equation}
For the same reason as above, the Poisson brackets in Eqs.~(\ref{eq59}) and 
(\ref{eq60}) vanish, and these equations can be brought into the following 
form:
\begin{equation}
  \tilde{I}_> \left( k \right) = \frac{\lambda}{6} \left( \tilde{G}_> \ast \tilde{G}_> \right) \left( k \right) \cdot \frac{1 - \tilde{I}_R \left( k \right)}{1 + \frac{\lambda}{3} \left( \coco{\tilde{G}_R} \ast \tilde{F} \right) \left( k \right)} \;,
\end{equation}
\begin{equation}
  \tilde{I}_< \left( k \right) = \frac{\lambda}{6} \left( \tilde{G}_< \ast \tilde{G}_< \right) \left( k \right) \cdot \frac{1 - \tilde{I}_R \left( k \right)}{1 + \frac{\lambda}{3} \left( \coco{\tilde{G}_R} \ast \tilde{F} \right) \left( k \right)} \;.
\end{equation}
With the aid of these equations we can show that in equilibrium the proper
self energy satisfies
\[ \tilde{\Sigma}_< \left( k \right) = \exp \left( - \beta k^0 \right) \tilde{\Sigma}_> \left( k \right) \;, \]
and together with the Fourier transformed Kubo-Martin-Schwinger condition 
(\ref{eq93}) for the propagator we see 
then immediately, that in equilibrium the collision term in the kinetic 
equation for the symmetric propagator indeed vanishes and that the kinetic 
equations really describe the equilibrium consistently.

\section{\label{sect3}Generalized Boltzmann Equation}
\setcounter{equation}{0}

The first step to learn something about the kinetic equations that we derived
in the last section is to consider the special case of a system containing
long-living single-particle excitations, where the width of the 
spectral function is much smaller than the effective particle mass. This 
allows us to employ a so-called quasi-particle approximation, and we can 
derive a generalized Boltzmann equation from the kinetic equations 
(\ref{eq31}) to (\ref{eq61}). Our starting point is the so-called 
Kadanoff-Baym ansatz, which is suggested by Eqs.~(\ref{eq94}) and 
(\ref{eq95}), namely:
\[ \tilde{G}_< \left( X, k \right) = \tilde{\varrho} \left( X, k \right) \tilde{n} \left( X, k \right) \;, \]
\[ \tilde{G}_> \left( X, k \right) = \tilde{\varrho} \left(X, k \right) \Big( 1 + \tilde{n} \left( X, k \right) \Big) \;. \]
It follows then that
\begin{equation} \label{eq28}
  \tilde{F} \left( X, k \right) = \tilde{\varrho} \left(X, k \right) \left( \tilde{n} \left( X, k \right) + \frac{1}{2} \right) \;.
\end{equation}
To employ a quasi-particle approximation means to assume that the particle 
number density does not depend explicitly on the energy, i.e.
\begin{equation} \label{eq54}
  \partial_{k^0} \tilde{n} \left( X, k \right) = 0 \;,
\end{equation}
and that the spectral function $\tilde{\varrho} \left( X, k \right)$ has the 
same form as for a free theory:
\begin{eqnarray*}
        \tilde{\varrho} \left( X, k \right) 
  & = & 2 \pi \cdot \sign \left( k^0 \right) \cdot \delta \left( {k^0}^2 - E^2 \left( X, \bm{k} \right) \right) \\
  & = & \frac{\pi}{E \left( X, \bm{k} \right)} \Big( \delta \left( k^0 - E \left( X, \bm{k} \right) \right) - \delta \left( k^0 + E \left( X, \bm{k} \right) \right) \Big) \;,
\end{eqnarray*}
where
\[ E \left( X, \bm{k} \right) = \sqrt{\bm{k}^2 + M^2 \left( X \right)} \;. \]
We see that we obtain this form of the spectral function from 
Eq.~(\ref{eq35}) when we set $\tilde{\Gamma} \left( X, k \right) = 
- \frac{1}{2} \tilde{\Sigma}_{\varrho} \left( X, k \right)$ and consider the 
limit $\tilde{\Gamma} \left( X, k \right) \to 0$. However, we have to 
emphasize the fact that we cannot send $\tilde{\Sigma}_{\varrho} \left( X, k 
\right) \to 0$ everywhere, as for example in the collision term, where the 
spectral self energy is necessary for thermalization. One can verify directly 
that in the sense of distributions
\begin{equation} \label{eq29} 
  \left( 2 k^{\mu} \partial_{X^{\mu}} + \left( \partial_{X^{\mu}} M^2 \left( X \right) \right) \partial_{k_{\mu}} \right) \tilde{\varrho} \left( X, k \right) = 0 \;.
\end{equation}
An immediate consequence from Eq.~(\ref{eq29}) is the fact that we can employ
the quasi-particle approximation consistently only, if the Poisson 
brackets on the right hand side of Eq.~(\ref{eq30}) vanish. The terms 
that contain derivatives with respect to $k^0$ vanish according to 
Eq.~(\ref{eq54}), and later we will consider spatially homogeneous systems, 
where also the terms with spatial derivatives vanish. However, for the moment 
we just simply assume that all Poisson brackets vanish, although we consider a 
spatially inhomogeneous system. Inserting Eqs.~(\ref{eq28}) and (\ref{eq29}), 
the left hand side of Eq.~(\ref{eq31}) becomes
\begin{eqnarray}
        LHS(\ref{eq31}) 
  & = & \frac{\pi}{E \left( X, \bm{k} \right)} \left( \delta \left( k^0 - E \left( X, \bm{k} \right) \right) - \delta \left( k^0 + E \left( X, \bm{k} \right) \right) \right) \qquad \label{eq32} \\
  &   & \qquad {} \times \left( 2 k^{\mu} \partial_{X^{\mu}} + \left( \partial_{X^{\mu}} M^2 \left( X \right) \right) \partial_{k_{\mu}} \right) \tilde{n} \left( X, k \right) \nonumber \;.
\end{eqnarray}
As we neglect the Poisson brackets, due to Eq.~(\ref{eq20}) and the 
Kadanoff-Baym ansatz the collision term on the right hand side reads:
\begin{eqnarray}
        RHS(\ref{eq31})
  & = & \frac{\pi}{E \left( X, \bm{k} \right)} \Big( \delta \left( k^0 - E \left( X, \bm{k} \right) \right) - \delta \left( k^0 + E \left( X, \bm{k} \right) \right) \Big) \label{eq33} \\
  &   & \qquad {} \times \left( \tilde{\Sigma}_> \left( X, k \right) \tilde{n} \left( X, k \right) - \tilde{\Sigma}_< \left( X, k \right) \left( 1 + \tilde{n} \left( X, k \right) \right) \right) \nonumber \;.
\end{eqnarray}
On the mass shell we define
\[ n \left( X, \bm{k} \right) \equiv \tilde{n} \left( X, \bm{k}, E \left( X, \bm{k} \right) \right) \;, \] 
which can be interpreted as a phase space distribution function for 
quasi-particles with momentum $\bm{k}$ and energy $E \left( X, \bm{k} 
\right)$. Equating the positive energy components of Eqs.~(\ref{eq32}) and 
(\ref{eq33}) and integrating over $k^0$ gives\footnote{This is the step where
it is necessary that the Poisson brackets in the kinetic equation for the 
symmetric propagator vanish. If this were not the case, we could not isolate 
the positive energy components on the right hand side.}:
\begin{eqnarray}
        \lefteqn{\left( \partial_{X^0} + \frac{k^j}{E \left( X, \bm{k} \right)} \partial_{X^j} + \frac{1}{2 E \left( X, \bm{k} \right)} \left( \partial_{X^j} M^2 \left( X \right) \right) \partial_{k_j} \right) n \left( X, \bm{k} \right)} \; \nonumber \\
  & = & \frac{1}{2 E \left( X, \bm{k} \right)} \Big( \Sigma_> \left( X, \bm{k} \right) n \left( X, \bm{k} \right) - \Sigma_< \left( X, \bm{k} \right) \left( 1 + n \left( X, \bm{k} \right) \right) \Big) \;. \qquad \label{eq79}
\end{eqnarray}
When we use the notation
\[ v^j = \frac{k^j}{E \left( X, \bm{k} \right)} \]
and the fact that
\[ \frac{1}{2 E \left( X, \bm{k} \right)} \left( \partial_{X^j} M^2 \left( X \right) \right) \partial_{k_j} = - \delta^{jl} \left( \partial_{X^j} E \left( X, \bm{k} \right) \right) \partial_{k^l} \;, \]
we can simplify the left hand side. On the right hand side we use the Wigner
transforms of Eqs.~(\ref{eq75}) and (\ref{eq76}) together with the identities
\[ \tilde{n} \left( X, -k \right) = - 1 - \tilde{n} \left( X, k \right) \;, \]
\[ \int \ddd{p} \left[ f \left( \bm{p} \right) \right] = \int \ddd{p} \left[ f \left( - \bm{p} \right) \right] \;, \]
such that we find a preliminary version of the generalized Boltzmann equation:
\begin{eqnarray}
        \lefteqn{\Big( \partial_{X^0} + \bm{v} \cdot \nabla_{\bm{X}} - \left( \nabla_{\bm{X}} E \left( X, \bm{k} \right) \right) \cdot \nabla_{\bm{k}} \Big) n \left( X, \bm{k} \right)} \; \label{eq36} \\
  & = & - \frac{\lambda}{12 N E \left( X, \bm{k} \right)} \int \frac{\ddd{p}}{\left( 2 \pi \right)^3} \Bigg[ \frac{1}{E \left( X, \bm{p} \right)} \nonumber \\
  &   & {} \times \Bigg\{ \tilde{I}_> \left( X, \bm{k} - \bm{p}, E \left( X, \bm{k} \right) - E \left( X, \bm{p} \right) \right) \left( 1 + n \left( X, \bm{p} \right) \right) n \left( X, \bm{k} \right) \nonumber \\
  &   & {} + \tilde{I}_> \left( X, \bm{k} - \bm{p}, E \left( X, \bm{k} \right) + E \left( X, \bm{p} \right) \right) n \left( X, \bm{p} \right) n \left( X, \bm{k} \right) \nonumber \\
  &   & {} - \tilde{I}_< \left( X, \bm{k} - \bm{p}, E \left( X, \bm{k} \right) - E \left( X, \bm{p} \right) \right) n \left( X, \bm{p} \right) \left( 1 + n \left( X, \bm{k} \right) \right) \nonumber \\
  &   & {} - \tilde{I}_< \left( X, \bm{k} - \bm{p}, E \left( X, \bm{k} \right) + E \left( X, \bm{p} \right) \right) \left( 1 + n \left( X, \bm{p} \right) \right) \left( 1 + n \left( X, \bm{k} \right) \right) \Bigg\} \Bigg] \nonumber \;.
\end{eqnarray}
This equation resembles very much the classical Boltzmann equation as it is 
presented for example in Ref.~\cite{La10}. However, there are some subtle 
differences that will be explained later. The comparison between the 
left hand side of this equation and Eq.~(3.3) in Ref.~\cite{La10} shows that 
$E \left( X, \bm{k} \right)$ acts as a potential for the quasi-particles. 

Now, after having pointed out the relation between the left hand sides of the
kinetic equation for the symmetric propagator and the classical Boltzmann 
equation, we specialize to spatially homogeneous systems, 
i.e. all quantities lose their dependence on the spatial center coordinate 
$\bm{X}$. As pointed out above, an immediate consequence of this is the fact 
that in this case all Poisson brackets vanish automatically. To simplify 
notation, we write $t$ instead of $X^0$ from now on. The resummation 
functions in Eq.~(\ref{eq36}) are related to the ones we used in the last 
section by
\[ \tilde{I}_> \left( t, k \right) = \tilde{I}_F \left( t, k \right) + \frac{1}{2} \tilde{I}_{\varrho} \left( t, k \right) \]
and 
\[ \tilde{I}_< \left( t, k \right) = \tilde{I}_F \left( t, k \right) - \frac{1}{2} \tilde{I}_{\varrho} \left( t, k \right) \;. \]
Hence, we obtain from Eqs.~(\ref{eq59}) and (\ref{eq60})
\begin{equation} \label{eq67}
  \tilde{I}_> \left( t, k \right) = \frac{\lambda}{6} \left( \tilde{G}_> \ast \tilde{G}_> \right) \left( t, k \right) \cdot \frac{1 - \tilde{I}_R \left( t, k \right)}{1 + \frac{\lambda}{3} \left( \coco{\tilde{G}_R} \ast \tilde{F} \right) \left( t, k \right)}
\end{equation}
and
\begin{equation} \label{eq68}
  \tilde{I}_< \left( t, k \right) = \frac{\lambda}{6} \left( \tilde{G}_< \ast \tilde{G}_< \right) \left( t, k \right) \cdot \frac{1 - \tilde{I}_R \left( t, k \right)}{1 + \frac{\lambda}{3} \left( \coco{\tilde{G}_R} \ast \tilde{F} \right) \left( t, k \right)} \;.
\end{equation}
We see that the resummation functions are given by the product of their 
contributions to the corresponding self energies at two-loop level (setting 
sun, which means three-loop level for $\Gamma_2$) times some factor which is 
the same for both resummation functions. This common factor includes the 
corrections coming from the resummation shown in Fig.~\ref{fig6}. Hence, we
can introduce an effective coupling associated with the bulky vertex shown
in Fig.~\ref{fig4}:
\begin{equation} \label{eq112}
  \effCo \left( t, k \right) = \frac{\lambda \left( 1 - \tilde{I}_R \left( t, k \right) \right)}{1 + \frac{\lambda}{3} \left( \coco{\tilde{G}_R} \ast \tilde{F} \right) \left( t, k \right)} \;.
\end{equation}
We will evaluate this effective coupling in the next section. The 
contributions of the resummation functions to the two-loop self energies
are given by
\begin{eqnarray}
        \lefteqn{\left( G_> \ast G_> \right) \left( t, k \right)} \; \label{eq55} \\
  & = & \frac{1}{2} \int \frac{\ddd{p}}{\left( 2 \pi \right)^3} \int \ddd{q} \Bigg[ \delta \left( \bm{q} - \left( \bm{k} - \bm{p} \right) \right) \frac{\pi}{E \left( t, \bm{q} \right) E \left( t, \bm{p} \right)} \nonumber \\
  &   & \times \Bigg\{ \delta \left( k^0 - \left( E \left( t, \bm{p} \right) + E \left( t, \bm{q} \right) \right) \right) \left( 1 + n \left( t, \bm{q} \right) \right) \left( 1 + n \left( t, \bm{p} \right) \right) \nonumber \\
  &   & {} + \delta \left( k^0 - \left( - E \left( t, \bm{p} \right) + E \left( t, \bm{q} \right) \right) \right) \left( 1 + n \left( t, \bm{q} \right) \right) n \left( t, \bm{p} \right) \nonumber \\
  &   & {} + \delta \left( k^0 - \left( E \left( t, \bm{p} \right) - E \left( t, \bm{q} \right) \right) \right) n \left( t, \bm{q} \right) \left( 1 + n \left( t, \bm{p} \right) \right) \nonumber \\
  &   & {} + \delta \left( k^0 - \left( - E \left( t, \bm{p} \right) - E \left( t, \bm{q} \right) \right) \right) n \left( t, \bm{q} \right) n \left( t, \bm{p} \right) \Bigg\} \Bigg] \nonumber
\end{eqnarray}
and
\begin{eqnarray}
        \lefteqn{\left( G_< \ast G_< \right) \left( t, k \right)} \; \label{eq56} \\
  & = & \frac{1}{2} \int \frac{\ddd{p}}{\left( 2 \pi \right)^3} \int \ddd{q} \Bigg[ \delta \left( \bm{q} - \left( \bm{k} - \bm{p} \right) \right) \frac{\pi}{E \left( t, \bm{q} \right) E \left( t, \bm{p} \right)} \nonumber \\
  &   & \times \Bigg\{ \delta \left( k^0 - \left( E \left( t, \bm{p} \right) + E \left( t, \bm{q} \right) \right) \right) n \left( t, \bm{q} \right) n \left( t, \bm{p} \right) \nonumber \\
  &   & {} + \delta \left( k^0 - \left( - E \left( t, \bm{p} \right) + E \left( t, \bm{q} \right) \right) \right) n \left( t, \bm{q} \right) \left( 1 + n \left( t, \bm{p} \right) \right) \nonumber \\
  &   & {} + \delta \left( k^0 - \left( E \left( t, \bm{p} \right) - E \left( t, \bm{q} \right) \right) \right) \left( 1 + n \left( t, \bm{q} \right) \right) n \left( t, \bm{p} \right) \nonumber \\
  &   & {} + \delta \left( k^0 - \left( - E \left( t, \bm{p} \right) - E \left( t, \bm{q} \right) \right) \right) \left( 1 + n \left( t, \bm{q} \right) \right) \left( 1 + n \left( t, \bm{p} \right) \right) \! \! \Bigg\} \Bigg] \nonumber \;.
\end{eqnarray}
When we insert Eqs.~(\ref{eq67}) to (\ref{eq56}) into Eq.~(\ref{eq36}), we 
obtain on the right hand side of Eq.~(\ref{eq36}) an integral over $\bm{p}$
and $\bm{q}$ of a sum of products which contain the energy $\delta$ 
functions appearing in Eqs.~(\ref{eq55}) and (\ref{eq56}), the effective 
coupling appearing in Eqs.~(\ref{eq67}) and (\ref{eq68}), and the $n$'s 
and $\left( 1 + n \right)$'s of Eqs.~(\ref{eq36}), (\ref{eq55}) and 
(\ref{eq56}). These products describe the various processes that can occur. 
$n$ represents an incoming particle and $\left( 1 + n \right)$ an outgoing 
one. Depending on the composition of the products, one can classify them to 
belong either to the gain terms or to the loss terms. For example the terms
\begin{equation} \label{eq57}
  n \left( t, \bm{k} \right) n \left( t, \bm{p} \right) \left( 1 + n \left( t, \bm{q} \right) \right) \left( 1 + n \left( t, \bm{k} - \bm{p} - \bm{q} \right) \right)
\end{equation}
and
\begin{equation} \label{eq58}
  \left( 1 + n \left( t, \bm{k} \right) \right) \left( 1 + n \left( t, \bm{p} \right) \right) n \left( t, \bm{q} \right) n \left( t, \bm{k} - \bm{p} - \bm{q} \right)
\end{equation}
describe the scattering of two particles. The term (\ref{eq57}) describes the 
scattering of a particle from the mode $\bm{k}$ at some particle from the mode
$\bm{p}$ into some other mode. Therefore, and because of its emerging with a 
minus sign, it belongs to the loss terms. On the other hand, the term 
(\ref{eq58}) describes the scattering of a particle from an arbitrary mode 
into the mode $\bm{k}$. Coming with a plus sign, this term belongs to the gain 
terms. Both processes are illustrated in 
\begin{figure}[t]
  \centering
  \includegraphics{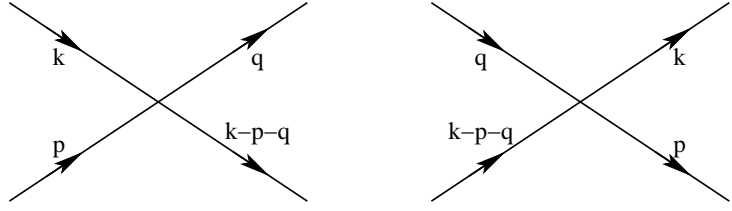}
  \caption{\label{fig2}Scattering of two particles.}
\end{figure}
Fig.~\ref{fig2}\footnote{If one finds the arrangement of the momenta 
inappropriate, because it contradicts the natural intuition of
momentum conservation, one can use that $n$ and $E$ depend only on the 
magnitude of their momentum arguments and perform a transformation for the 
momenta, to obtain the natural arrangement.}.

\begin{figure}[t]
  \centering
  \includegraphics{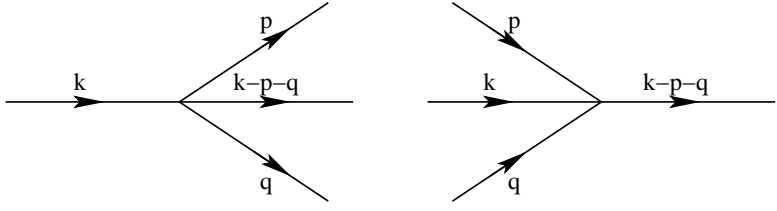}
  \caption{\label{fig3}Decay and recombination processes.}
\end{figure}

Similarly, the following two terms describe --- and Fig.~\ref{fig3} 
illustrates --- the decay of one particle into three particles and the 
recombination of three particles to one particle, respectively:
\begin{equation}
  n \left( t, \bm{k} \right) \left( 1 + n \left( t, \bm{p} \right) \right) \left( 1 + n \left( t, \bm{q} \right) \right) \left( 1 + n \left( t, \bm{k} - \bm{p} - \bm{q} \right) \right)
\end{equation}
and
\begin{equation}
  n \left( t, \bm{k} \right) n \left( t, \bm{p} \right) n \left( t, \bm{q} \right) \left( 1 + n \left( t, \bm{k} - \bm{p} - \bm{q} \right) \right) \;.
\end{equation}
However, momentum conservation prevents the arguments of the respective energy 
$\delta$ functions from being zero. Hence, these decay and recombination 
processes are forbidden and we can cancel these terms in our equation.
We prove this statement for one special case. The energy satisfies the 
following inequalities:
\[ E \left( t, \bm{k} + \bm{p} \right) < \sqrt{4 M^2 \left( t \right) + \left( \bm{k} + \bm{p} \right)^2} \le E \left( t, \bm{k} \right) + E \left( t, \bm{p} \right) \;. \]
The first inequality results from the fact that in the symmetric regime the 
effective mass is always strictly greater than zero and the second one is the 
triangle inequality for the Euclidean norm in $\R^4$ applied to the 
vectors $\left( M 
\left( t \right), \bm{k} \right)$ and $ \left( M \left( t \right), \bm{p} 
\right)$. The replacement $\bm{k} \to \bm{k} - \bm{p}$ reveals the following 
complementary inequality:
\[ E \left( t, \bm{k} \right) - E \left( t, \bm{p} \right) < E \left( t, \bm{k} - \bm{p} \right) \;. \]
An appropriate choice of the momenta then shows that
\[ E \left( t, \bm{k} \right) - E \left( t, \bm{p} \right) < E \left( t, \bm{k} - \bm{p} \right) < E \left( t, \bm{q} \right) + E \left( t, \bm{k} - \bm{p} - \bm{q} \right) \]
\[ \Longleftrightarrow \qquad E \left( t, \bm{k} \right) - E \left( t, \bm{p} \right) - E \left( t, \bm{q} \right) - E \left( t, \bm{k} - \bm{p} - \bm{q} \right) < 0 \]
Hence, the term with the $\delta$ function that has the left hand side of the 
last inequality as its argument vanishes. With very much the same reasoning 
one can rule out all the remaining decay and recombination terms.

Furthermore, it is obvious that terms which describe four incoming particles 
leaving to nowhere or four outgoing particles coming from nowhere vanish, 
because the arguments of the respective energy $\delta$ functions have no 
zeros.

Eventually, we obtain for the generalized Boltzmann equation\footnote{We use 
the following abbreviations: $E_{\bm{k}} \equiv E \left( t, \bm{k} \right)$ 
and $n_{\bm{k}} \equiv n \left( t, \bm{k} \right)$}:
\begin{myeqnarray}{B1}
        \lefteqn{\partial_t n \left( t, \bm{k} \right) = - \frac{\lambda \pi}{144 N} \int \frac{\ddd{p}}{\left( 2 \pi \right)^3} \int \frac{\ddd{q}}{\left( 2 \pi \right)^3} \Bigg[ \frac{1}{E_{\bm{k}} E_{\bm{p}} E_{\bm{q}} E_{\bm{k} - \bm{p} - \bm{q}}}} \; \label{eq64} \\
  &   & {} \times \left( 2 \effCo \Big( t, \bm{k} - \bm{q}, E_{\bm{k}} - E_{\bm{q}} \Big) + \effCo \Big( t, \bm{k} - \bm{p}, E_{\bm{k}} + E_{\bm{p}} \Big) \right) \nonumber \\
  &   & {} \times \delta \Big( E_{\bm{k}} + E_{\bm{p}} - E_{\bm{q}} - E_{\bm{k} - \bm{p} - \bm{q}} \Big) \nonumber \\
  &   & {} \times \Big( n_{\bm{k}} n_{\bm{p}} \left( 1 + n_{\bm{q}} \right) \left( 1 + n_{\bm{k}-\bm{p}-\bm{q}} \right) - \left( 1 + n_{\bm{k}} \right) \left( 1 + n_{\bm{p}} \right) n_{\bm{q}} n_{\bm{k}-\bm{p}-\bm{q}} \Big) \Bigg] \nonumber \;.
\end{myeqnarray}In 
contrast to the classical Boltzmann equation our generalized Boltzmann 
equation includes an effective mass and an effective coupling. The effective
mass, which in Eq.~(\ref{eq64}) is hidden in the energies, arises from the
local tadpole contributions to the self energy (cf.~Fig.~\ref{fig4}) and 
satisfies a gap equation as will be shown in the next section. The effective
coupling arises from the resummation of the infinite series of diagrams that
contribute to the nonlocal part of the self energy (cf.~Figs.~\ref{fig4} and
\ref{fig6}). As we will show in the next section this resummation takes higher
products of $n$'s and $(1+n)$'s into account. Therefore, in contrast to the
classical Boltzmann equation, we expect our generalized Boltzmann equation to 
be valid even for large particle number densities, where one cannot neglect 
collisions between three or more particles.

Before we finish this section, we illuminate briefly our expectations about 
the solution of Eq.~(\ref{eq64}). As our system contains only bosonic 
quasi-particles, the equilibrium particle number density is the 
Bose-Einstein distribution function
\begin{equation} \label{eq66}
  n_B \left( \bm{k} \right) = \frac{1}{\exp \left( \frac{E \left( \bm{k} \right)}{T} \right) - 1} \;.
\end{equation}
In equilibrium there is no time-dependence at all, and 
\[ E \left( \bm{k} \right) = \sqrt{M_{eq}^2 + \bm{k}^2} \;. \]
The temperature $T$ of the system is defined only in equilibrium and is 
determined by fitting the numerical data that is obtained at the end of the 
numerical evolution of the Boltzmann equation to the Bose-Einstein 
distribution. In equilibrium the effective mass is a constant and is denoted 
by $M_{eq}$. Furthermore, after the system has equilibrated, the particle 
number density must not change anymore. Consequently, the collision integral 
on the right hand side of the Boltzmann equation must vanish. We show that 
this is indeed the case by simply inserting the Bose-Einstein distribution
function (\ref{eq66}) into the collision integral on the right hand side of 
the generalized Boltzmann equation (\ref{eq64})\footnote{As in equilibrium 
there is no time dependence at all, here we use the abbreviation 
$E_{\bm{k}} \equiv E \left( \bm{k} \right)$. $\beta$ is as usual the inverse
temperature.}:
\begin{eqnarray*}
        \lefteqn{\int \frac{\ddd{p}}{\left( 2 \pi \right)^3} \int \frac{\ddd{q}}{\left( 2 \pi \right)^3} \Bigg[ \frac{\delta \Big( E_{\bm{k}} + E_{\bm{p}} - E_{\bm{q}} - E_{\bm{k} - \bm{p} - \bm{q}} \Big)}{E_{\bm{k}} E_{\bm{p}} E_{\bm{q}} E_{\bm{k}-\bm{p}-\bm{q}}}} \; \\
  &   & {} \times \frac{2 \effCo \Big( t, \bm{k} - \bm{q}, E_{\bm{k}} - E_{\bm{q}} \Big) + \effCo \Big( t, \bm{k} - \bm{p}, E_{\bm{k}} + E_{\bm{p}} \Big)}{\left( e^{\beta E_{\bm{k}}} - 1 \right) \left( e^{\beta E_{\bm{p}}} - 1 \right) \left( e^{\beta E_{\bm{q}}} - 1 \right) \left( e^{\beta E_{\bm{k}-\bm{p}-\bm{q}}} - 1 \right)} \\
  &   & {} \times \Big( \exp \left( \beta E_{\bm{q}} \right) \exp \left( \beta E_{\bm{k}-\bm{p}-\bm{q}} \right) - \exp \left( \beta E_{\bm{k}} \right) \exp \left( \beta E_{\bm{p}} \right) \Big) \Bigg] \\
  & = & \int \frac{\ddd{p}}{\left( 2 \pi \right)^3} \int \frac{\ddd{q}}{\left( 2 \pi \right)^3} \Bigg[ \frac{\delta \Big( E_{\bm{k}} + E_{\bm{p}} - E_{\bm{q}} - E_{\bm{k} - \bm{p} - \bm{q}} \Big)}{E_{\bm{k}} E_{\bm{p}} E_{\bm{q}} E_{\bm{k}-\bm{p}-\bm{q}}} \\
  &   & {} \times \frac{2 \effCo \Big( t, \bm{k} - \bm{q}, E_{\bm{k}} - E_{\bm{q}} \Big) + \effCo \Big( t, \bm{k} - \bm{p}, E_{\bm{k}} + E_{\bm{p}} \Big)}{\left( e^{\beta E_{\bm{k}}} - 1 \right) \left( e^{\beta E_{\bm{p}}} - 1 \right) \left( e^{\beta E_{\bm{q}}} - 1 \right) \left( e^{\beta E_{\bm{k}-\bm{p}-\bm{q}}} - 1 \right)} \\
\end{eqnarray*}
\begin{eqnarray*}
  &   & {} \times \Big( \exp \left( \beta \left( E_{\bm{k}} + E_{\bm{p}} - E_{\bm{k}-\bm{p}-\bm{q}} \right) \right) \exp \left( \beta E_{\bm{k}-\bm{p}-\bm{q}} \right) - \exp \left( \beta \left( E_{\bm{k}} + E_{\bm{p}} \right) \right) \Big) \Bigg] \\
  & = & 0
\end{eqnarray*}
As was expected, our generalized Boltzmann equation contains a consistent 
description of the equilibrium. In the next section we will derive equations
that determine the effective mass and the effective coupling for any given
particle number density at some time $t$.

\section{\label{sect4}Effective Mass and Effective Coupling}
\setcounter{equation}{0}

\subsection*{Effective Mass}

According to Eq.~(\ref{eq74}) the effective mass is given by
\[ M^2 \left( X \right) = m^2 + \lambda \frac{N + 2}{6 N} G \left( X, X \right) \;. \]
We use that 
\[ G \left( X, X \right) = F \left( X, X \right) = \tilde{F} \left( X, 0 \right) \]
and express the symmetric propagator in terms of its Fourier transform. The 
quasi-particle approximation then leads to a gap equation for the effective 
mass, which has to be solved self-consistently.
\begin{myeqnarray}{B2}
        M^2 \left( X \right) 
  & = & m^2 + \lambda \frac{N + 2}{6 N} \int \frac{\dddd{p}}{\left( 2 \pi \right)^4} \left[ \tilde{F} \left( X, p \right) \right] \nonumber \\
  & = & m^2 + \lambda \frac{N + 2}{12 N} \int \frac{\ddd{p}}{\left( 2 \pi \right)^3} \left[ \frac{2 n \left( X, \bm{p} \right) + 1}{\sqrt{M^2 \left( X \right) + \bm{p}^2}} \right] 
\end{myeqnarray}Of 
course, for a spatially homogeneous system the effective mass loses its 
dependence on $\bm{X}$, and $M^2 \left( X \right)$ becomes $M^2 \left( t 
\right)$.

\subsection*{Effective Coupling}

Taking into account that Eq.~(\ref{eq61}) yields
\begin{myequation}{B3} \label{eq103} 
  \tilde{I}_R \left( t, k \right) = \frac{\lambda}{3} \cdot \frac{\left( \tilde{G}_R \ast \tilde{F} \right) \left( t, k \right)}{1 + \frac{\lambda}{3} \left( \tilde{G}_R \ast \tilde{F} \right) \left( t, k \right)} \;,
\end{myequation}we 
obtain for the effective coupling (\ref{eq112})
\begin{myequation}{B4} \label{eq104}
  \effCo \left( t, k \right) = \frac{\lambda}{\left| 1 + \frac{\lambda}{3} \left( \tilde{G}_R \ast \tilde{F} \right) \left( t, k \right) \right|^2} \;.
\end{myequation}The 
convolution in the denominator yields
\begin{eqnarray*}
        \left( \tilde{G}_R \ast \tilde{F} \right) \left( X, k \right) 
  & = & \frac{1}{2} \int \frac{\ddd{p}}{\left( 2 \pi \right)^3} \Bigg[ \frac{n \left( t, \bm{p} \right) + \frac{1}{2}}{E \left( t, \bm{p} \right)} \Big( \tilde{G}_R \left( t, \bm{k} - \bm{p}, k^0 - E \left( t, \bm{p} \right) \right) \\
  &   & {} + \tilde{G}_R \left( t, \bm{k} - \bm{p}, k^0 + E \left( t, \bm{p} \right) \right) \Big) \Bigg]
\end{eqnarray*}
where the retarded propagator is given by Eq.~(\ref{eq24}) as
\begin{myequation}{B5} \label{eq105}
  \tilde{G}_R \left( t, k \right) = \frac{1}{- k^2 + M^2 \left( t \right) + \tilde{\Sigma}_R \left( t, k \right)} \;.
\end{myequation}Due 
to Eqs.~(\ref{eq106}) and (\ref{eq59}) the expression for the retarded 
self energy reads
\begin{myeqnarray}{B6}
        \tilde{\Sigma}_R \left( t, k \right)
  & = & - \frac{\lambda}{6 N} \int \frac{\ddd{p}}{\left( 2 \pi \right)^3} \Bigg[ \frac{n \left( t, \bm{p} \right) + \frac{1}{2}}{E \left( t, \bm{p} \right)} \Big( \tilde{I}_R \left( t, \bm{k} - \bm{p}, k^0 - E \left( t, \bm{p} \right) \right) \nonumber \\
  &   & \qquad {} + \tilde{I}_R \left( t, \bm{k} - \bm{p}, k^0 + E \left( t, \bm{p} \right) \right) \Big) \Bigg] \label{eq107} \\
  &   & {} - \frac{\lambda}{72 N} \int \frac{\ddd{p}}{\left( 2 \pi \right)^3} \int \frac{\ddd{q}}{\left( 2 \pi \right)^3} \Bigg[ \frac{1}{E \left( t, \bm{p} \right) E \left( t, \bm{q} \right)} \nonumber \\
  &   & {} \times \Bigg\{ \tilde{G}_R \left( t, \bm{k} - \bm{p} - \bm{q}, k^0 - E_{\bm{p}} - E_{\bm{q}} \right) \effCo \left( t, \bm{p} + \bm{q}, E_{\bm{p}} + E_{\bm{q}} \right) \nonumber \\
  &   & \qquad {} \times \left( \left( n \left( t, \bm{p} \right) + \frac{1}{2} \right) \left( n \left( t, \bm{q} \right) + \frac{1}{2} \right) + \frac{1}{4} \right) \nonumber \\
  &   & {} + 2 \tilde{G}_R \left( t, \bm{k} - \bm{p} - \bm{q}, k^0 - E_{\bm{p}} + E_{\bm{q}} \right) \effCo \left( t, \bm{p} + \bm{q}, E_{\bm{p}} - E_{\bm{q}} \right) \nonumber \\
  &   & \qquad {} \times \left( - \left( n \left( t, \bm{p} \right) + \frac{1}{2} \right) \left( n \left( t, \bm{q} \right) + \frac{1}{2} \right) + \frac{1}{4} \right) \nonumber \\
  &   & {} + \tilde{G}_R \left( t, \bm{k} - \bm{p} - \bm{q}, k^0 + E_{\bm{p}} + E_{\bm{q}} \right) \effCo \left( t, \bm{p} + \bm{q}, - E_{\bm{p}} - E_{\bm{q}} \right) \nonumber \\
  &   & \qquad {} \times \left( \left( n \left( t, \bm{p} \right) + \frac{1}{2} \right) \left( n \left( t, \bm{q} \right) + \frac{1}{2} \right) + \frac{1}{4} \right) \Bigg\} \Bigg] \;. \nonumber
\end{myeqnarray}We 
find that Eqs.~(\ref{eq64}) to (\ref{eq107}) form a closed set of 
equations which determine the evolution of the particle number density 
completely. In the next section we are going to present methods which allow
for the tracing of this evolution for a given initial particle number 
density.

\section{Numerical Implementation}
\setcounter{equation}{0}

In order to solve the kinetic equations for the symmetric propagator and the 
spectral function or the generalized Boltzmann equation numerically, we 
propose to use a standard lattice discretization \cite{MoMue}. We consider 
our whole system of physical particles to live on a lattice enclosed in a 
spatial cube with periodic boundary conditions. When doing so, the cube and 
the periodic boundary conditions cause the momenta to be discrete. Conversely, 
because of the lattice the momenta are periodic, which means that the lattice 
takes care of the regularization. Physically reasonable results are obtained 
by approaching the continuum as well as the infinite volume limit.

The edges of the cube have length $L$, such that its volume is $V = L^3$. The 
lattice spacing is $a_s$, and $N_s$ is the number of lattice bins along each 
edge, such that 
\[ L = N_s a_s \;. \]
For a spatially homogeneous system the lattice discretization indicated above
yields for the momenta:
\[ \bm{k}^2 \to \sum_{j=1}^3 \frac{4}{a_s^2} \sin^2 \left( \frac{a_s k_{n_j}}{2} \right) \;, \]
where as usual, the momenta $\kk$ are discretized according to
\[ \kk = \frac{2 \pi}{L} \bm{n} \;, \]
with
\[ \bm{n} \in \left\{ 0, \ldots, N_s-1 \right\}^3 \;.\]
As a consequence, we also have to reverse the thermodynamic limit and perform 
the following replacement:
\[ \int \frac{\ddd{p}}{\left( 2 \pi \right)^3} \to \frac{1}{V} \sum_{\bm{n} \in \left\{ 0, \ldots, N_s-1 \right\}^3} \;. \]

\subsection*{Generalized Boltzmann Equation}

With the aid of these means the discretized version of the generalized
Boltzmann equation (\ref{eq64}) reads
\begin{eqnarray*}
        \lefteqn{\partial_t n \left( t, \kk \right) = - \frac{\lambda \pi}{144 N V^2} \quad \sum_{\bm{m}, \bm{l}} \Bigg[ \frac{1}{E_{\kk} E_{\pp} E_{\qq} E_{\kk-\pp-\qq}}} \; \\
  &   & {} \times \left( 2 \effCo \Big( t, \kk - \qq, E_{\kk} - E_{\qq} \Big) + \effCo \Big( t, \kk - \pp, E_{\kk} + E_{\pp} \Big) \right) \\
  &   & {} \times \delta \Big( E_{\kk} + E_{\pp} - E_{\qq} - E_{\kk - \pp - \qq} \Big) \\
  &   & {} \times \Big( n_{\kk} n_{\pp} \left( 1 + n_{\qq} \right) \left( 1 + n_{\kk-\pp-\qq} \right) \\
  &   & \qquad {} - \left( 1 + n_{\kk} \right) \left( 1 + n_{\pp} \right) n_{\qq} n_{\kk-\pp-\qq} \Big) \Bigg] \;.
\end{eqnarray*}
The discretized $\delta$ function has become a Kronecker $\delta$, i.e. it is 
1, if its argument vanishes and 0 otherwise. The discretized energy is given 
by
\[ E_{\pp} = \sqrt{M^2 \left( t \right) + \sum_{j} \frac{4}{a_s^2} \sin^2 \left( \frac{a_s p_{m_j}}{2} \right)} \]
and the effective mass satisfies
\begin{equation} \label{eq108}
  M^2 \left( t \right) = m^2 + \lambda \frac{N + 2}{12 N V} \sum_{\bm{m}} \frac{2 n \left( t, \pp \right) + 1}{\sqrt{M^2 \left( t \right) + \sum_{j} \frac{4}{a_s^2} \sin^2 \left( \frac{a_s p_{m_j}}{2} \right)}} \;.
\end{equation}
Equivalently, one can discretize the four equations that determine the 
effective coupling which were given in the previous section. 

We see that due to its character as an integro-differential equation the
discretized Boltzmann equation becomes a system of $N_s^3$ coupled 
ordinary differential equations. We recall that the generalized Boltzmann 
equation stems from the kinetic equations for the symmetric propagator and
the spectral function which were supposed to be valid in the 
intermediate-time drifting regime, where the evolution of the system is 
sufficiently smooth. Of course, this property is inherited by the generalized 
Boltzmann equation, which means that the particle number density changes only 
slightly on its approach to the Bose-Einstein distribution function.
Therefore we propose to use a fourth order Runge-Kutta method with adaptive 
stepsize control to solve the Boltzmann equation numerically \cite{NumRec}. 
Due to its adaptive stepsize control this method exploits the smooth behaviour 
of the particle number density and will increase the stepsize very rapidly 
which allows for an efficient approach to the equilibrium.

The gap equation (\ref{eq108}) for the effective mass has to be solved after
each time step self-consistently. This can be done with a Newton-Raphson 
method \cite{NumRec}.
Unfortunately this method is not applicable to solve Eqs.~(\ref{eq103}) to 
(\ref{eq107}) for the coupling. Here, one is restricted to an iterative 
approach to $\effCo$.

Examining the collision sum on the right hand side of the discretized 
Boltzmann equation more closely, we find that the complexity to compute
this sum explicitely is of order $\mathcal{O} \left( N_s^6 \right)$. 
Accordingly, this would make the numerical solution of the generalized 
Boltzmann equation an order $\mathcal{O} \left( N_s^9 \right)$ problem, which 
in practice
makes the approach to the infinite volume limit impossible. To reduce the 
complexity of the computation of the collision sum one therefore has to use
stochastic (Monte-Carlo) methods to estimate its value. Such stochastic
methods were originally invented to compute higher dimensional integrals by
exploiting the law of large numbers, namely that for a function $f$
defined in a higher dimensional volume $V$ and $R$ random variables 
$x_1, \ldots, x_R \in V$, we have
\[ \frac{1}{R} \sum_{j=1}^R f \left( x_j \right) \quad \xrightarrow{R \to \infty} \quad \overline{f} \equiv \frac{1}{V} \intl_V \dd{x} \left[ f \left( x \right) \right] \;. \]
Therefore we can estimate the integral by
\[ \intl_V \dd{x} \left[ f \left( x \right) \right] = \frac{V}{R} \sum_{j=1}^R f \left( x_j \right) \;, \]
where the error in the estimation is of order $\mathcal{O}(1 / \sqrt{R})$. 
It is straightforward to apply this estimation to the collision sum.
For randomly chosen $m_1, m_2, m_3, l_1, l_2$ one performs the sum over $l_3$
to eliminate the energy Kronecker $\delta$. As we still have $N_s^3$ equations
but only one sum left, we have reduced the complexity for solving the 
generalized Boltzmann equation to order $\mathcal{O} \left( N_s^4 \right)$.
Nevertheless, we would like to mention that in giving only the order of the 
complexity we have neglected the constants. Especially $R$ may become
very large (between $10^3$ and $10^5$) such that for small volumes there is 
only little increase in the efficiency of our numerics. For example, for 
$N_s = 32$ and $R = 10^4$ one saves a factor of about $3 \cdot 10^3$ for 
every collision sum that has to be computed.

In order to solve the Boltzmann equation, one starts with an initial particle 
number density, i.e. for each $\kk$ one specifies a certain value $n \left( 0, 
\kk \right)$. For this given density, one then has to compute $n \left( t_1, 
\kk \right)$ for every momentum mode $\kk$ after the first timestep, thus
obtaining the particle number density for this time. This procedure has to be
repeated until the particle number density does not change anymore up to some
given accuracy, such that one can assume that the system has reached its 
equilibrium state. The temperature of the system can then be obtained by 
fitting the particle number density to the Bose-Einstein distribution 
function.

\subsection*{Kinetic Equations}

It is our aim to trace the evolution of the propagator for some given 
far-from-equilibrium initial conditions with the aid of the full evolution
equations (\ref{eq50}) to (\ref{eq87}) until we find that every momentum mode
has entered the intermediate-time drifting regime. Having arrived at this 
point, we would like to switch to the kinetic equations in order to observe
thermalization in a more efficient way. The first step, namely the numerical
implementation to solve the full evolution equations for a spatially 
homogeneous system, is described in detail in Ref.~\cite{Be1}. The remainder 
of this section is devoted to the description of the second and the third 
step, namely the transition from the full evolution equations to the kinetic 
equations as well as the numerical implementation to solve the latter ones. 
These two steps also are described for a spatially homogeneous system.

Consider Fig.~\ref{fig7} in the introduction and suppose that for some time 
$t_0$ all momentum modes have entered the intermediate-time drifting regime.
For example for $t=100$, we see that this is the case for at least all 
momentum modes shown in Fig.~\ref{fig7}. Then we could choose $t_0$ to be the
time for the transition to the kinetic equations. Of course, the larger $t_0$
the more accurately the kinetic equations describe the subsequent approach to
thermal equilibrium. The transition itself is done in the following way: 
Taking the effective memory loss shown in Fig.~\ref{fig8} into account, one
can introduce a cut-off at the lower limits of the memory integrals:
\[ \intl_0^{x^0} \dd{z^0} \longrightarrow \intl_{x^0-S}^{x^0} \dd{z^0} \;. \]
$S$ denotes the size of the storage where we save the values of the propagator
for previous times. Let $a_t$ be the stepsize in time direction and $N_t$ the
number of values stored for each quantity at previous times, then
\[ S = N_t a_t \;. \]
In order to perform the transition at the time $t_0$ we have to compute the 
evolution of the propagator with the full evolution equations up to 
times $x^0 = y^0 = t_0 + (S/2)$, where we have stored the values of the 
propagator for all times
\[ \left( x^0, y^0 \right) \in \left\{ t_0 - \frac{S}{2}, t_0 - \frac{S}{2} + a_t, \ldots, t_0 + \frac{S}{2} \right\}^2 \;. \]
For fixed center time $t_0$ we see that this set contains $N_t+1$ pairs
$\left( x^0, y^0 \right)$ with different values for $s^0$. Hence, the
energy dependence of the Wigner transformed propagator is also
encoded in $N_t+1$ different values of the energy and we can perform the Wigner 
transformation according to
\begin{eqnarray*}
        F \left( x^0, y^0, \kk \right) 
  & \longrightarrow & \tilde{F} \left( t_0, s^0_m, \kk \right) \\
  & \longrightarrow & \tilde{F} \left( t_0, \kk, \omega_m \right) = a_t \sum_{j=0}^{N_t} \exp \left( i \omega_m s^0_j \right) \tilde{F} \left( t_0, \kk, s^0_j \right) \;,
\end{eqnarray*}
where $m \in \left\{ 0, \ldots, N_t \right\}$ and
\[ \omega_m = \frac{2 \pi}{S} m \;. \]
This transformation has to be performed for every spatial momentum mode 
$\kk$ and every of the functions $F$, $\varrho$, $\Sigma_F$, 
$\Sigma_{\varrho}$, $I_F$, $I_{\varrho}$, $G_R$, $\Sigma_R$ and $I_R$.
Especially for the retarded quantities it is crucial to use the above 
procedure in order to avoid the convolutions with the Fourier transformed 
$\theta$ function.

After having performed the transition from the full evolution equations to 
the kinetic equations, one can then use the standard lattice discretization
described at the beginning of this section to discretize the kinetic 
equations and solve them numerically using standard techniques. Exactly as 
for the generalized Boltzmann equation, we suggest to use a fourth order 
Runge-Kutta method with adaptive stepsize control to advance in time.
Exploiting the smooth behaviour of the subsequent evolution of the propagator 
this method will increase the stepsize rapidly and thus allow for an efficient 
approach to thermal equilibrium.

\chapter*{Conclusions and Outlook}
\markboth{\em Conclusions and Outlook}{\em Conclusions and Outlook}
\addcontentsline{toc}{chapter}{Conclusions and Outlook}
\thispagestyle{empty}

Starting from the $1/N$-expansion of the 2PI effective action to 
next-to-leading order, we derived kinetic equations for a systematic 
nonperturbative description of the evolution of $O(N)$-symmetric real scalar 
quantum fields out of thermal equilibrium.

The techniques presented here have several generic advantages. First of all,
evolution equations derived from the effective action are certain to be 
time-reversal invariant, and the corresponding kinetic equations are known 
to possess exact rather than approximate conservation laws on the level of 
expectation values.
Additionally, the $1/N$-expansion provides a nonperturbative description of 
quantum fields out of equilibrium. As a consequence the range of 
applicability of our kinetic equations is not limited by the question of the 
validity of a loop expansion, but instead is restricted only by the gradient 
expansion employed in their derivation. In situations where this 
gradient expansion is justified, our kinetic equations describe the evolution 
of quantum fields out of equilibrium without further approximations. In 
particular we showed that our kinetic equations remain valid in the case of 
nonperturbatively large fluctuations.

Furthermore, we showed how one can get from our kinetic equations to a 
generalized Boltzmann equation for quasi-particles. While the three-loop 
approximation of the 2PI effective action comprises the classical Boltzmann 
equation \cite{CaHu1,BlIa}, the $1/N$-expansion of the 2PI effective action 
to next-to-leading order leads to a generalized Boltzmann equation which 
includes an effective mass and an effective coupling.

Finally, this diploma thesis can serve as a starting point for further 
research. In very much the same way as we obtained our kinetic equations in 
this work, one should be able to derive kinetic equations in the case of 
broken symmetries where the field expectation value does not vanish. This 
would be important to describe the phenomenon of parametric resonance for 
large times. Apart from that, as already indicated in the introduction, our 
kinetic equations are expected to be valid near a second-order phase 
transition. Additionally,
we mentioned several times that we expect our generalized Boltzmann equation
to be valid even for large particle number densities, where one cannot neglect
collisions between three or more particles. Of course, these expectations have 
to be tested. Further aims are the application of these methods to fermionic 
and gauge theories.

\chapter*{Acknowledgements}
\thispagestyle{empty}
There are quite a lot of people I would like to thank very much: \\[1cm]
First of all, I would like to thank Dr.~J\"urgen Berges and Prof.~Christof 
Wetterich for giving me the possibility to carry out my diploma thesis on this 
very interesting and active subject, for many interesting and fruitful 
discussions and for their guidance and patience throughout the last 
year.\\[0.5cm]
I thank Prof. Dieter Gromes for taking over the co-correction.\\[0.5cm]
Furthermore, I spent a very nice time in the ``Westzimmer'' with Thomas 
``Ach\-tung! Tipp mal ein!'' Auer, Christian ``Lord of the dark\ldots'' 
M\"uller (The dots stand either for `ness' or for ' energy'. The reader might 
judge himself.) and Felix Schwab. The latter one has no nickname. He is 
standing above the image problem. Our discussions on physics, politics, books, 
sports and movies were very enlightening.\\[0.5cm]
I thank Christian and Felix for proof-reading this work.\\[0.5cm]
Last but not least, I would like to thank my parents for everything they did
for me throughout the last 25 years.

\clearpage
\thispagestyle{empty}
\cleardoublepage
\thispagestyle{empty}

\noindent\underline{\large\bf Erkl\"arung:} \\[1cm]

\noindent Ich versichere, da\3 ich diese Arbeit selbst\"andig verfa\3t und 
keine anderen als die angegebenen Hilfsmittel benutzt habe. \\[0.5cm]
Heidelberg, den \dotfill \hspace{4.5cm} \dotfill\\
\hspace*{\fill} Markus M. M\"uller

\end{document}